\documentclass[aps,prd,twocolumn,superscriptaddress,
preprintnumbers,showpacs,nofootinbib,notitlepage,
floatfix,10pt]{revtex4-2}
\usepackage{graphicx}
\usepackage{subfigure}
\usepackage{amsmath}
\usepackage{amssymb}
\usepackage{slashed}
\usepackage{amsfonts}
\usepackage{xcolor}
\usepackage{multirow}
\usepackage{array}
\usepackage{mathrsfs}
\usepackage{bbm}
\usepackage{comment}
\usepackage{MnSymbol}

\usepackage[colorlinks=true]{hyperref}

\usepackage{multirow}

\begin{document}
\title{Lattice QCD calculation of charmed baryon decay constants at continuum limit and physical mass}%

\author{Lei-Yi Li}
\affiliation{State Key Laboratory of Dark Matter Physics, Key Laboratory for Particle Astrophysics and Cosmology (MOE),
Shanghai Key Laboratory for Particle Physics and Cosmology,
School of Physics and Astronomy, Shanghai Jiao Tong University, Shanghai 200240, China}

\author{Jie Ran}
\affiliation{State Key Laboratory of Dark Matter Physics, Key Laboratory for Particle Astrophysics and Cosmology (MOE),
Shanghai Key Laboratory for Particle Physics and Cosmology,
School of Physics and Astronomy, Shanghai Jiao Tong University, Shanghai 200240, China}
 
\author{Meng-Chu Cai}
\affiliation{Institute of Theoretical Physics, Chinese Academy of Sciences, Beijing 100190, China}

\author{Hao-Fei Gao}
\affiliation{State Key Laboratory of Dark Matter Physics, Key Laboratory for Particle Astrophysics and Cosmology (MOE), Shanghai Key Laboratory for Particle Physics and Cosmology, School of Physics and Astronomy, Shanghai Jiao Tong University, Shanghai 200240, China}
\affiliation{Tsung-Dao Lee Institute, Shanghai Jiao Tong University, Shanghai 201210, China}

\author{Yu Gu}
\affiliation{Department of Physics, College of Physics and Optoelectronic Engineering,
Jinan University, Guangzhou 510632, P.R. China}

\author{Xue-Ying Han}
\affiliation{Institute of High Energy Physics, Chinese Academy of Sciences, Beijing 100049, China}
\affiliation{School of Physical Sciences, University of Chinese Academy of Sciences, Beijing 100049, China}

\author{Jun Hua}
\affiliation{Key Laboratory of Atomic and Subatomic Structure and Quantum Control (MOE), 
Guangdong Basic Research Center of Excellence for Structure and Fundamental Interactions of Matter, 
Institute of Quantum Matter, South China Normal University, Guangzhou 510006, China}
\affiliation{Guangdong-Hong Kong Joint Laboratory of Quantum Matter, 
Guangdong Provincial Key Laboratory of Nuclear Science, Southern Nuclear Science Computing Center, 
South China Normal University, Guangzhou 510006, China}

\author{Jin-Xin Tan}
\affiliation{State Key Laboratory of Dark Matter Physics, Key Laboratory for Particle Astrophysics and Cosmology (MOE), Shanghai Key Laboratory for Particle Physics and Cosmology, School of Physics and Astronomy, Shanghai Jiao Tong University, Shanghai 200240, China}
\affiliation{Tsung-Dao Lee Institute, Shanghai Jiao Tong University, Shanghai 201210, China}

\author{Guang-Yu Wang}
\email[Corresponding author: ]{guangyuwang@sjtu.edu.cn}
\affiliation{State Key Laboratory of Dark Matter Physics, Key Laboratory for Particle Astrophysics and Cosmology (MOE),
Shanghai Key Laboratory for Particle Physics and Cosmology,
School of Physics and Astronomy, Shanghai Jiao Tong University, Shanghai 200240, China}

\author{Wei Wang}
\email[Corresponding author: ]{wei.wang@sjtu.edu.cn}
\affiliation{State Key Laboratory of Dark Matter Physics, Key Laboratory for Particle Astrophysics and Cosmology (MOE),
Shanghai Key Laboratory for Particle Physics and Cosmology,
School of Physics and Astronomy, Shanghai Jiao Tong University, Shanghai 200240, China}
 
\author{Fanrong Xu}
\affiliation{Department of Physics, College of Physics and Optoelectronic Engineering,
Jinan University, Guangzhou 510632, P.R. China}

\author{Yi-Bo Yang}
\affiliation{University of Chinese Academy of Sciences, School of Physical Sciences, Beijing 100049, China}
\affiliation{Institute of Theoretical Physics, Chinese Academy of Sciences, Beijing 100190, China}
\affiliation{School of Fundamental Physics and Mathematical Sciences, Hangzhou Institute for Advanced Study, UCAS, Hangzhou 310024, China}
\affiliation{International Centre for Theoretical Physics Asia-Pacific, Beijing/Hangzhou, China}

\author{Qi-An Zhang}
\email[Corresponding author: ]{zhangqa@buaa.edu.cn}
\affiliation{School of Physics, Beihang University, Beijing 102206, China}

\date{December 2025}%

\begin{abstract}
We present the first principle calculation of charmed baryon decay constants employing 2+1 flavor gauge ensembles with lattice spacings ranging from 0.05 to 0.1 fm and pion masses between 136 and 310 MeV. Under $SU(3)$ flavor symmetry, we construct the charmed baryon interpolating operators and compute the corresponding hadronic matrix elements to extract the bare decay constants for each ensemble. The non-perturbative renormalization is performed via the symmetric momentum-subtraction scheme. After performing systematic chiral and continuum extrapolations, we obtain the decay constants with a precision of $8\sim 16\%$ from first principles.
\end{abstract}

\maketitle

\clearpage
\twocolumngrid

\section{Introduction}

\begin{table*}
    \centering
    \renewcommand{\arraystretch}{2.0}
  \setlength{\tabcolsep}{2.5mm}
    \caption{Details of the parameters used for the calculations on each ensembles. The number of gauge configurations are denoted as $N_{\mathrm{cfg}}$, and for each configuration, $N_{\mathrm{meas}}$ measurements are performed.}
    \begin{tabular}{ccccccccc}
    \toprule 
    Ensemble & $L^{3}\times T$ & $a$(fm) & $\tilde{m}_{l}^{{b}}$ & $\tilde{m}_{s}^{{v}}$ & $\tilde{m}_{c}^{{v}}$ & $m_{\pi}$(MeV) & $N_{\mathrm{cfg}}$ &  $N_{\mathrm{meas}}$ \\
    \hline
    C24P29 & $24^{3}\times72$ & 0.10524 & -0.2770 & -0.2356 & 0.4159 & 292.3  & 864 & $20$ \\
    C32P29 & $32^{3}\times64$ & 0.10524 & -0.2770 & -0.2358 & 0.4150 & 293.1  & 984 & $20$ \\
    C32P23 & $32^{3}\times64$ & 0.10524 & -0.2790 & -0.2337 & 0.4190 & 227.9  & 452 & $20$ \\
    C48P14 & $48^{3}\times96$ & 0.10524 & -0.2825 & -0.2335 & 0.4205 & 136.4  & 187 & $40$ \\
    F32P30 & $32^{3}\times96$ & 0.07753 & -0.2295 & -0.2038 & 0.1974 & 300.4  & 777 & $20$ \\
    H48P32 & $48^{3}\times144$ & 0.05199 & -0.1850 & -0.1701 & 0.0551 & 316.6 & 550 & $12$ \\
    \toprule 
    \end{tabular}
    \label{tab:CLQCD}
\end{table*}

Charmed baryons provide a unique laboratory for exploring the interplay between heavy-quark dynamics and nonperturbative QCD, and their decay constants are among the most fundamental nonperturbative parameters characterizing these systems. In particular, the decay constants of ground-state charmed baryons enter the theoretical description of both leptonic and semileptonic processes, where they control the overlap between the heavy quark and the light diquark degrees of freedom and thus directly affect the normalization of decay amplitudes. Precise knowledge of these quantities is essential for reliable extractions of Cabibbo-Kobayashi-Maskawa (CKM) matrix elements from charm decays, for testing heavy-quark symmetry and its breaking patterns, and for constraining models of hadron structure that are routinely used to interpret data from LHCb, Belle II, and future flavor facilities. At the same time, charmed baryon decay constants offer a clean benchmark for nonperturbative methods such as lattice QCD, QCD sum rules, and effective field theory approaches, because they are short-distance-safe and highly sensitive to the internal structure of the baryon. A systematic and precise determination of these decay constants is therefore a crucial step toward reducing theoretical uncertainties in charm physics and deepening our understanding of baryonic bound states in QCD.

In recent years, driven by the accumulation of high-precision experimental data, the study of charmed baryons has entered an era of precision measurements. The uncertainty of the charmed-baryon decay constant directly limits the accuracy of theoretical predictions \cite{Wang:2009cr, Wang:2010fq, Zhao:2020mod, Shi:2019hbf, Khodjamirian:2011jp, Miao:2022bga, Huber:2016xod, Zhang:2022iun, Rui:2024xgc, Rui:2025iwa}, and thus affects indirect searches for possible signals of new physics. A precise determination of these decay constants is therefore indispensable. Lattice QCD provides a first-principles framework for computing nonperturbative quantities, with the advantage of being model independent and systematically improvable, thereby ensuring reliable theoretical control. The decay constants of charmed mesons have already been computed with high precision in previous lattice study \cite{Li:2024vtx, CLQCD:2024yyn, Follana:2007uv, Davies:2010ip, PACS-CS:2011ngu, FermilabLattice:2011njy, Na:2012iu,  Dimopoulos:2013qfa, Yang:2014sea, Carrasco:2014poa, Carrasco:2014poa,  Boyle:2017jwu, Bazavov:2017lyh, Bazavov:2017lyh, Chen:2020qma, Dimopoulos:2021qsf, Kuberski:2024pms, FlavourLatticeAveragingGroupFLAG:2024oxs}. In contrast, for charmed baryons the available results currently come only from QCD sum rules \cite{Wang:2009cr, Wang:2010fq, Zhao:2020mod, Shi:2019hbf, Khodjamirian:2011jp}. This makes a first-principles lattice determination of charmed-baryon decay constants both timely and necessary.

In this work, we present a lattice QCD determination of the charmed-baryon decay constants based on $2+1$ flavor gauge ensembles generated by the CLQCD Collaboration \cite{CLQCD:2024yyn}, using the tadpole-improved tree-level Clover fermion action together with the tadpole-improved Symanzik gauge action \cite{CLQCD:2023sdb}. These ensembles, which employ stout-smeared Clover fermions \cite{Liu:2022gxf}, were generated at three lattice spacings, $a = (0.05199,0.07753,0.10524)~\text{fm}$, thereby allowing for a controlled continuum extrapolation. The ensembles cover a range of pion masses, including the physical point, which allows a controlled chiral extrapolation. The basic parameters of all ensembles used in this work are summarized in Tab.~\ref{tab:CLQCD}, and further details can be found in Refs.~\cite{CLQCD:2024yyn, CLQCD:2023sdb}.

On these ensembles, the bare decay constants of charmed baryons are extracted from two-point correlation functions constructed with suitably chosen interpolating operators. In this work, we build charmed-baryon interpolating operators based on $SU(3)$ flavor symmetry, which provide good overlap with the ground states of the target baryons. For each ensemble we compute the corresponding two-point functions using multiple source positions, and determine the ground-state masses and amplitudes from fits to the large-time behavior of the correlation functions. The bare decay constants are then obtained from these ground-state amplitudes according to the appropriate continuum normalization conventions.

To convert the bare decay constants into physical quantities, we renormalize the charmed-baryon operators nonperturbatively. The renormalization constants are determined in the symmetric momentum-subtraction (SMOM) scheme and subsequently converted to the $\overline{\mathrm{MS}}$ scheme using perturbative conversion factors. The renormalized decay constants for all ensembles are finally quoted at the common scale $\mu = 2~\text{GeV}$. By performing combined chiral and continuum extrapolations of the renormalized results, we obtain charmed-baryon decay constants with total uncertainties in the range of approximately $8\sim16\%$, providing the first lattice-QCD determination of these quantities from first principles.

The rest of this paper is organized as follows. In Sec.~\ref{sec:2}, we introduce the interpolating operators and two-point correlation functions for singly charmed baryons, and present the lattice determination of their masses and bare decay constants. In Sec.~\ref{sec:3}, we describe in detail the nonperturbative renormalization of the baryonic operators. The renormalized decay constants on each ensemble, as well as the combined results extrapolated to the continuum and physical (chiral) limits, are reported in Sec.~\ref{sec:4}. Finally, we summarize our findings and discuss possible future directions in Sec.~\ref{sec:5}.

\section{Interpolators and Correlators for Charmed Baryons}
\label{sec:2}

In the quark model, the singly charmed baryons with spin–parity $J^P=(1/2)^+$ are classified according to the $SU(3)$ flavor symmetry of the two light quarks. They form an antitriplet $\mathbf{\bar{3}}$ and a sextet $\mathbf{6}$. The antitriplet $\mathcal{B}_{c}^{\bar{3}}=(\Lambda_{c}^{+},\,\Xi_{c}^{+},\,\Xi_{c}^{0})$ arises from coupling the charm quark to a spin-zero, antisymmetric light diquark, while the sextet $\mathcal{B}_{c}^{6}=(\Sigma_{c}^{++},\, \Sigma_{c}^{+},\, \Sigma_{c}^{0},\, \Xi_{c}^{\prime+},\,\Xi_{c}^{\prime0},\,\Omega_{c}^{0})$ originates from a spin-one, symmetric light diquark. In the nonrelativistic limit, these diquark configurations reduce to the familiar pseudoscalar and vector diquark structures of the constituent quark model, and this flavor–spin classification persists in full QCD once the corresponding three-quark currents are properly renormalized.

The construction of suitable baryon currents provides the theoretical basis for defining the decay constants in QCD and for computing the two-point correlation functions from which they are extracted. In Ref.~\cite{Khodjamirian:2011jp}, interpolating currents for the $\Lambda_c$ and $\Sigma_c$ baryons were constructed using diquark $SU(2)$ symmetry, and the associated decay constants were evaluated within the framework of QCD sum rules. In Ref.~\cite{Liu:2023feb}, the authors generalized this construction to the $\Xi_c$ and $\Xi_c'$ baryons using $SU(3)$ flavor symmetry and determined the mixing angle between the corresponding interpolating fields in lattice QCD. In all these approaches, the decay constants are defined through matrix elements of local three-quark operators that carry the same quantum numbers as the physical baryons, QCD interactions then renormalize these currents but do not alter their underlying flavor–spin structure.

\begin{figure}
\centering
		\scalebox{0.43}[0.43]{\includegraphics[width=1\textwidth]{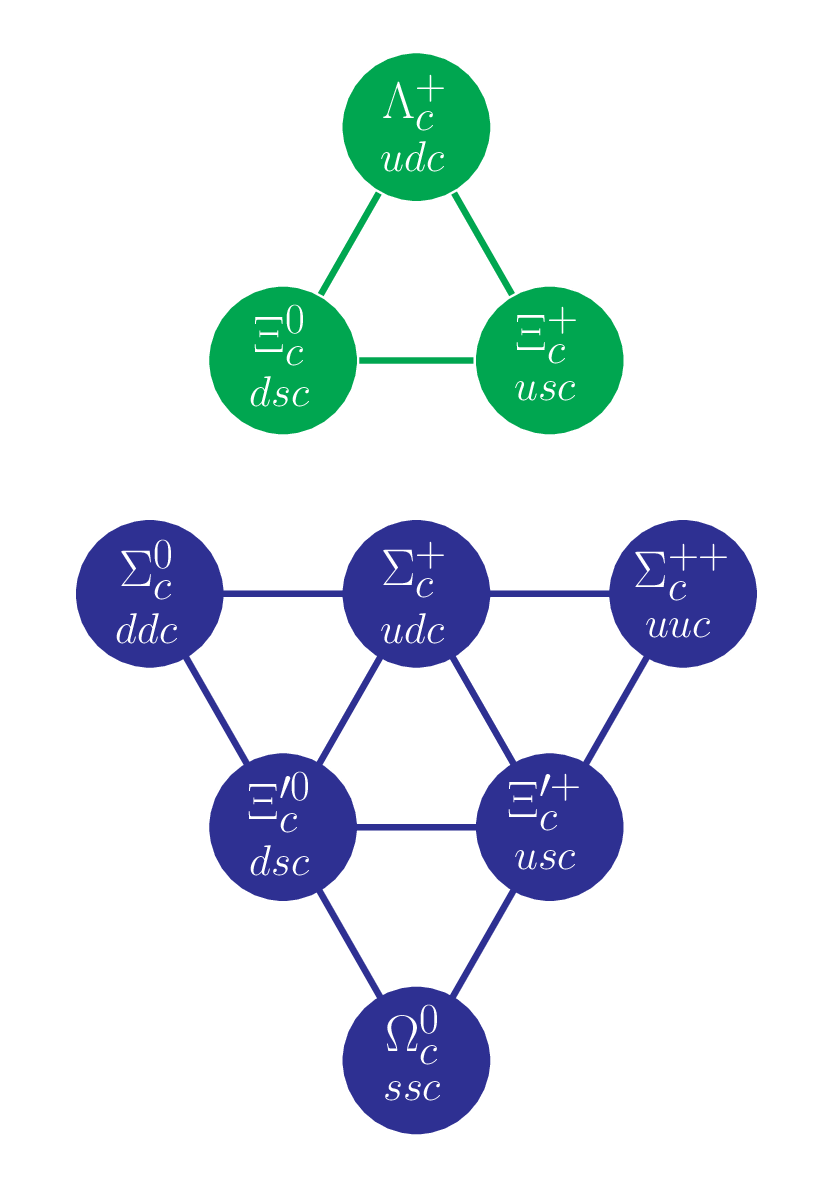}}
		\caption{$SU(3)$ flavor classification of the $J^{P}=(1/2)^{+}$ charmed baryons. Green triangles denote the anti-triplet states, while blue inverted triangles denote the sextet states.
 }
\label{fig: charmed baryon}
\end{figure}

In the present work, we adopt an $SU(3)$ flavor classification of the charmed baryons based on the decomposition $\mathbf{3 \otimes 3 = \bar{3} \oplus 6}$ for the two light quarks~\cite{He:2021qnc,Liu:2023feb}. The corresponding interpolating operators are used both to define the decay constants in the continuum and as source–sink currents in our lattice calculation. Fig.~\ref{fig: charmed baryon} illustrates the quark-flavor content and the $SU(3)$ symmetry classification of the singly charmed baryons. 

Guided by the diquark $SU(3)$ flavor symmetry, the antitriplet charmed-baryon operators are constructed from spin-zero light diquarks and are represented by pseudoscalar currents~\cite{Khodjamirian:2011jp,Liu:2023feb}:
\begin{equation}
\bar{3}:\left\{ \begin{aligned}O_{\Lambda_{c}^{+}} & =\epsilon_{ijk}(u_{i}^{T}C\gamma_{5}d_{j})c_{k},\\
O_{\Xi_{c}^{+}} & =\epsilon_{ijk}(u_{i}^{T}C\gamma_{5}s_{j})c_{k},\\
O_{\Xi_{c}^{0}} & =\epsilon_{ijk}(d_{i}^{T}C\gamma_{5}s_{j})c_{k},
\end{aligned}
\right.
\label{eq: charmed baryon anti-triplt}
\end{equation}
whereas the sextet operators are built from spin-one light diquarks and take the form of vector currents~\cite{Khodjamirian:2011jp,Liu:2023feb}:
\begin{equation}
    6:\left\{ \begin{aligned}O_{\Xi_{c}^{\prime+}} & =\epsilon_{ijk}(u_{i}^{T}C\gamma_{\mu}s_{j})\gamma^{\mu}\gamma_{5}c_{k},\\
O_{\Xi_{c}^{\prime0}} & =\epsilon_{ijk}(d_{i}^{T}C\gamma_{\mu}s_{j})\gamma^{\mu}\gamma_{5}c_{k},\\
O_{\Sigma_{c}^{+}} & =\epsilon_{ijk}(u_{i}^{T}C\gamma_{\mu}d_{j})\gamma^{\mu}\gamma_{5}c_{k},\\
O_{\Sigma_{c}^{0}} & =\frac{1}{\sqrt{2}}\epsilon_{ijk}(d_{i}^{T}C\gamma_{\mu}d_{j})\gamma^{\mu}\gamma_{5}c_{k},\\
O_{\Sigma_{c}^{++}} & =\dfrac{1}{\sqrt{2}}\epsilon_{ijk}(u_{i}^{T}C\gamma_{\mu}u_{j})\gamma^{\mu}\gamma_{5}c_{k},\\
O_{\Omega_{c}^{0}} & =\dfrac{1}{\sqrt{2}}\epsilon_{ijk}(s_{i}^{T}C\gamma_{\mu}s_{j})\gamma^{\mu}\gamma_{5}c_{k}.
\end{aligned}
\right.
\label{eq: charmed baryon sextet}
\end{equation}
Here $C$ denotes the charge-conjugation matrix, the Latin indices $i,~j,~k$ denote color, and $\epsilon_{ijk}$ is the totally antisymmetric tensor in color space. The factor $1/\sqrt{2}$ ensures the proper normalization of operators containing two identical light quarks. In the lattice calculation, these operators are used as interpolating currents in the two-point correlation functions; their ground-state overlap amplitudes, after appropriate renormalization, are matched to the continuum matrix elements that define the charmed-baryon decay constants.

The decay constant of a charmed baryon is defined through the matrix element
\begin{equation}
\begin{aligned}\langle0|O_{\mathcal{B}_{c}}|\mathcal{B}_{c}\rangle & =m_{\mathcal{B}_{c}}f_{\mathcal{B}_{c}}u_{\mathcal{B}_{c}},\end{aligned}
\label{eq: matrix element}
\end{equation}
where $m_{\mathcal{B}_{c}}$, $f_{\mathcal{B}_{c}}$, $u_{\mathcal{B}_{c}}$, and $O_{\mathcal{B}_{c}}$ denote the mass, bare decay constant, Dirac spinor, and interpolating operator of the charmed baryon, respectively. The construction of suitable operators thus provides the foundation for computing the two-point correlation functions from which the decay constants are extracted.

With the interpolating operators for charmed baryons established in this section, we now turn to the lattice determination of the bare decay constants. Throughout this work, a superscript $(0)$ is used to indicate bare quantities prior to renormalization. The Euclidean two-point correlation function for a charmed baryon $\mathcal{B}_{c}$ is defined as
\begin{equation}
    C_{\mathrm{2pt}}(t,\vec{p})=\sum_{\vec{x}}e^{-i\vec{p}\cdot\vec{x}}\langle O_{\mathcal{B}_{c}}^{(0)}(x),\bar{O}_{\mathcal{B}_{c}}^{(0)}(0)\rangle,
\end{equation}
where $x=(t,\vec{x})$ and $\bar{O}=O^{\dagger}\gamma^t$. In this work, we focus on the baryons at rest and hence set $\vec{p}=0$. To isolate the positive-parity contribution, we project with
\begin{equation}
    C_{\mathrm{2pt}}^{+}(t)=\mathrm{Tr}\left[C_{\mathrm{2pt}}(t,\vec{0})P_{+}\right],\qquad
P_{+} = \frac{\mathbbm{1}+\gamma_{t}}{2},
\end{equation}
so that $C_{\mathrm{2pt}}^{+}(t)$ predominantly receives contributions from states with quantum numbers $J^P=(1/2)^+$. 

\begin{figure}
\centering
		\hspace{0cm}\scalebox{0.4}[0.4]{\includegraphics[width=1\textwidth]{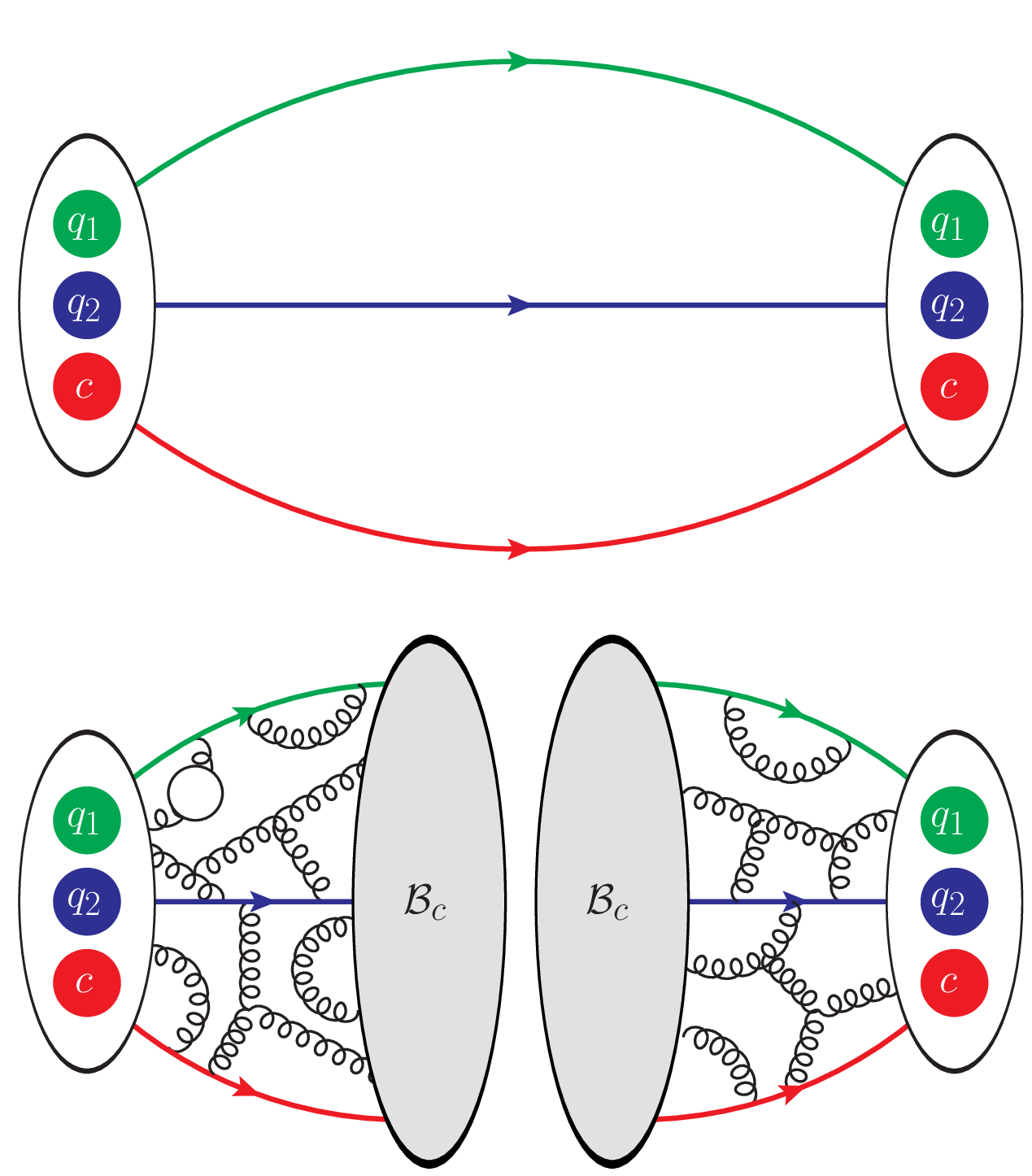}}
		\caption{The two-point correlation functions of charmed baryons, shown from top to bottom, correspond to the quark level and the hadron level, respectively.}
\label{fig:2pt}
\end{figure}

By inserting a complete set of hadronic states,
\begin{equation}
    \mathbbm{1}=\dfrac{1}{V}\sum_{n}\dfrac{1}{2E_{n}}|n\rangle\langle n|,
\end{equation}
where $V$ is the spatial volume, and using Euclidean time evolution, one obtains the standard spectral decomposition of the two-point function, as illustrated schematically in Fig.~\ref{fig:2pt}. At zero momentum and large Euclidean time separations, the correlator is dominated by the ground-state contribution of $\mathcal{B}_{c}$, with exponentially suppressed excited-state contaminations. Making use of the decay-constant definition in Eq.~\eqref{eq: matrix element} and the usual relativistic normalization of baryon states, the projected correlator can be parametrized as
\begin{equation}
    \begin{aligned}C_{\mathrm{2pt}}^{+}(t) & =2\left(m_{\mathcal{B}_{c}}f_{\mathcal{B}_{c}}^{(0)}\right)^2e^{-m_{\mathcal{B}_{c}}t}\left(1+de^{-\Delta m t} +\cdots \right),
    \end{aligned}
    \label{eq: C2pt}
\end{equation}
where $f_{\mathcal{B}_{c}}^{(0)}$ is the bare decay constant of the charmed baryon, $d$ denotes the fit parameter describing the contributions from higher excited states, and $\Delta m$ denotes the mass gap between the first excited state and the ground state. The ellipsis indicates further excited states, which are negligible in the chosen fit window.
In practice, we extract $m_{\mathcal{B}_{c}}$ and $f_{\mathcal{B}_{c}}^{(0)}$ by fitting Eq.~\eqref{eq: C2pt} to the lattice data for $C_{\mathrm{2pt}}^{+}(t)$ in a range of $t$ where the excited-state contamination is under control and the signal-to-noise ratio remains acceptable.

In our simulations, we employ point-source and point-sink two-point functions without additional spatial smearing of the baryon operators. This choice ensures that the overlap amplitude of the ground state in Eq.~\eqref{eq: C2pt} coincides directly with the matrix element in Eq.~\eqref{eq: matrix element}, and no extra normalization factors associated with smeared operators are introduced into the relation between the fitted amplitude and the bare decay constant. 
For each ensemble, we calculate the two-point functions using multiple source positions on each gauge configuration to enhance statistics. The numbers of configurations and measurements on each configuration used in this work are summarized in Tab.~\ref{tab:CLQCD}.

\begin{figure}
\centering
		\scalebox{0.43}[0.43]{\includegraphics[width=1\textwidth]{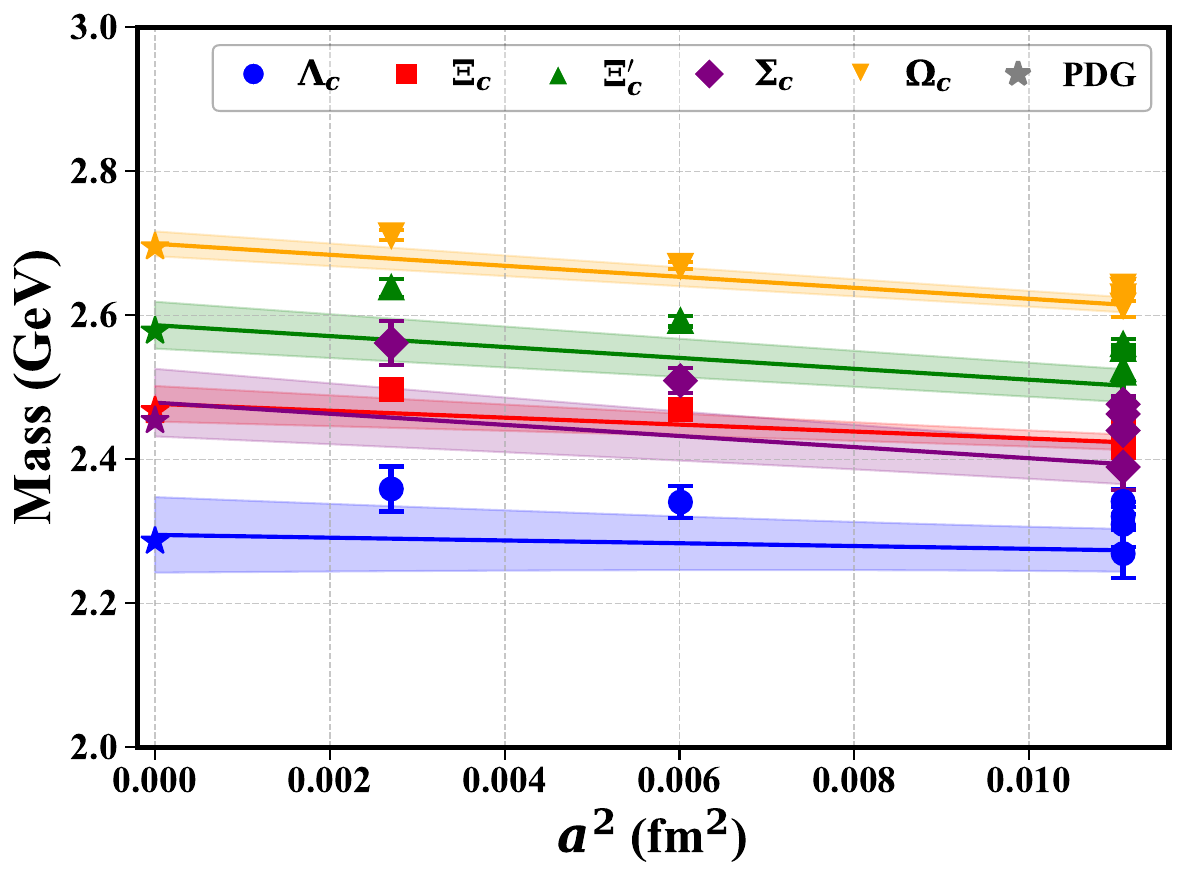}}
		\caption{The chiral and continuum extrapolation results for the charmed baryon masses are shown as a function of the lattice spacing. The blue, red, green, purple and yellow error bands correspond to $\Lambda_{c}$, $\Xi_{c}$, $\Xi^{\prime}_{c}$, $\Sigma_{c}$ and $\Omega_{c}$ baryons, respectively. }
\label{fig: mass}
\end{figure}

For each ensemble, high-statistics two-point correlators are obtained by computing on a large number of gauge configurations and measurements. To ensure a robust determination of the ground-state quantities, we analyze these correlators using several choices of fit ranges (and, where appropriate, different excited-state ans\"atze) and combine the corresponding results using a Bayesian model-averaging procedure~\cite{Jay:2020jkz}. This strategy reduces the dependence on any particular fit window and allows the fit-related systematics to be reflected in the quoted uncertainties, leading to stable and reliable values for $m_{\mathcal{B}_{c}}$ and $f_{\mathcal{B}_{c}}^{(0)}$. The individual ensemble results for $m_{\mathcal{B}_{c}}$ and their extrapolation to the physical point are displayed in Fig.~\ref{fig: mass}, and their agreement with the corresponding Particle Data Group (PDG) values \cite{ParticleDataGroup:2024cfk} provides a validation of our methodology and overall analysis. Further technical details of the fits and the model-averaging prescription are summarized in the Appendix. {A more systematic analysis of charmed baryon masses can be found in Ref.~\cite{Hu:2024mas}.}

\section{Renormalization for Bare Decay Constants}
\label{sec:3}

In the previous subsection, we have obtained the bare decay constants $f_{\mathcal{B}_{c}}^{(0)}$ associated with the charmed-baryon interpolating operators. Since these operators are composite, ultraviolet divergences inevitably arise in their matrix elements, and an appropriate renormalization scheme must be applied in order to obtain physically meaningful and comparable decay constants. The renormalization of baryon operators in lattice QCD has been extensively investigated in a variety of contexts, such as proton decay \cite{JLQCD:1999dld, Aoki:2006ib, Yoo:2021gql} and the calculation of baryon light-cone distribution amplitudes \cite{QCDSF:2008zfe, QCDSF:2008qtn, RQCD:2019hps}. In the present work, we adapt and extend the nonperturbative renormalization framework originally developed for proton decay operators \cite{Yoo:2021gql} to the case of charmed-baryon currents, and use it to determine the corresponding renormalization constants entering the definition of the decay constants.

The calculation of nonperturbative renormalization constants depends on the choice of renormalization scheme. Ref.~\cite{Aoki:2006ib,Yoo:2021gql} compute the renormalization of composite operators using the MOM three quark (MOM3q) and SMOM three quark (SYM3q) schemes, respectively. As shown in the comparison presented in \cite{Yoo:2021gql}, the SYM3q/$\mathrm{SMOM_{\gamma_{\mu}}}$ scheme exhibits better convergence of the conversion factors in MOM3q/MOM scheme. Therefore, in this work, we adopt the SYM3q/$\mathrm{SMOM_{\gamma_{\mu}}}$ renormalization scheme to compute the renormalization constants of the charmed baryon operators.
The SYM3q/$\mathrm{SMOM_{\gamma_{\mu}}}$ renormalization condition is defined as follows \cite{QCDSF:2008zfe}: 
    \begin{equation}    Z_{\mathcal{B}_{c}}Z_{q_{1}}^{-1/2}Z_{q_{2}}^{-1/2}Z_{q_{3}}^{-1/2}\Gamma_{\mathcal{B}_{c}}\big|_{\mathrm{SYM3q/SMOM_{\gamma_{\mu}}}}=1,
    \label{eq: charmed baryon renormalization condition origin}
\end{equation}
where $Z_{q_{i}}, i=1\sim3$ denotes the quark field renormalization constant, and $\Gamma_{\mathcal{B}_{c}}$ represents the projected vertex functions of charmed baryon, and $Z_{\mathcal{B}_{c}}$ is the renormalization constants of charmed baryon operators in Eq.~(\ref{eq: charmed baryon anti-triplt}) and (\ref{eq: charmed baryon sextet}). The definition of bare quantity and renormalization terms are:
\begin{equation}
\begin{aligned}
    q=&Z_{q}q^{(0)}_{q},\\
    O_{\mathcal{B}_{c}}=&Z_{\mathcal{B}_{c}}O_{\mathcal{B}_{c}}^{(0)}.
\end{aligned}
\end{equation}
In the calculation of renormalization constants, the ultraviolet corrections at the vertex are independent of the quark mass. Furthermore, since the perturbative conversion factor is derived in massless quark limit \cite{Gracey:2012gx,Yoo:2021gql}, the nonperturbative determination of the renormalization constant must also be performed with massless quarks. In this lattice calculation, we treat the $u,d,s,c$ quarks as light quarks \cite{Yoo:2021gql}, leading to a further simplification of the charmed baryon operators in Eq.~(\ref{eq: charmed baryon anti-triplt}) and (\ref{eq: charmed baryon sextet}) into the following form:
\begin{equation}
    \begin{aligned}
    O_{1} & =\epsilon_{ijk}(q_{1,i}^{T}C\gamma_{5}q_{2,j})q_{3,k},\\
    O_{2} & =\epsilon_{ijk}(q_{1,i}^{T}C\gamma_{\mu}q_{2,j})\gamma^{\mu}\gamma_{5}q_{3,k},\\
    O_{3} & =\epsilon_{ijk}(q_{1,i}^{T}C\gamma_{\mu}q_{1,j})\gamma^{\mu}\gamma_{5}q_{3,k},
\label{eq:O1-O3}
\end{aligned}
\end{equation}
where $O_{1}$ represents the operators of anti-triple baryon $\Lambda_{c}^{+},\, \Xi_{c}^{+},\, \Xi_{c}^{0}$ in light quark mass, and $O_{2}$ and $O_{3}$ represents the operators of sextet baryon $\Xi_{c}^{\prime +},\, \Xi_{c}^{\prime 0},\, \Sigma_{c}^{+}$ and $\Sigma_{c}^{++},\, \Sigma_{c}^{0},\, \Omega_{c}^{0}$ respectively. In massless quark limit, the renormalization condition simplifies to:
\begin{equation}
    Z_{O_{i}}Z_{l}^{-3/2}\Gamma_{O_{i}}\big|_{\mathrm{SYM3q/SMOM_{\gamma_{\mu}}}}=1,
    \label{eq: charmed baryon renormalization condition}
\end{equation}
where $Z_{l}$ is the renormalization constant of the light quark field. $Z_{O_{i}}$ are the renormalization constants of operators $O_{1}\sim O_{3}$. {The projected vertex functions $\Gamma_{\mathcal{B}_{c}}$ in Eq.~(\ref{eq: charmed baryon renormalization condition origin}) are replaced by $\Gamma_{O_{i}}$ and the definition are given in Eq.~(\ref{eq: Gamma_Oi}).}

 \subsection{Projected vertex functions of bilinear operators}
In the previous subsection, we presented the renormalization conditions for the charmed baryon operators in Eq.~(\ref{eq: charmed baryon renormalization condition}). It is clear that the calculation requires the use of the quark field renormalization constant, $Z_l$. Ref.~\cite{Yoo:2021gql} eliminates the quark field renormalization constant $Z_l$ by calculating the renormalization relation of the axial-vector current operator.
 In this work, the calculation of the renormalization constants is carried out using the configurations generated by the clover action. Therefore, we eliminate the quark field renormalization constant $Z_l$ by using the renormalization condition of vector current operator $O^{\mu}_{V}=\bar{q}\gamma^{\mu}q$ in $\mathrm{SMOM_{\gamma_{\mu}}}$ scheme:
\begin{equation}
    Z_{V}Z_{l}^{-1}\Gamma_{V}\big|_{\mathrm{SMOM_{\gamma_{\mu}}} }=1.
\label{eq: bilinear renormalization condition}
\end{equation}
The projected vertex function for the vector current operator $\Gamma_{V}$ is defined as:
\begin{equation}
    \Gamma_{V}=\dfrac{1}{48}\mathrm{Tr}[\Lambda_{V}^{\mu}(p_{1},p_{2})\gamma_{\mu}],
\end{equation}
where $\Lambda_{V}^{\mu}(p_{1},p_{2})$ is the amputated Green's function, and it satisfies the following definition:
\begin{equation}
    \Lambda_{V}^{\mu}(p_{1},p_{2})=S^{-1}(p_{1})G_{V}^{\mu}(p_{1},p_{2})S^{-1}(p_{2}).
\end{equation}
In the above equation, $S(p) = \sum_{x} e^{-ip \cdot x} \langle q(0) \bar{q}(x) \rangle$ represents the propagator. The definition of the Green's function in momentum space $G_{V}^{\mu}(p_{1},p_{2})$ is given by:
\begin{equation}
    G_{V}^{\mu}(p_{1},p_{2}) =\sum_{x,y}e^{-i(p_{1}\cdot x -p_{2}\cdot y)}\langle q(x)O^{\mu}(0)\bar{q}(y)\rangle.
\end{equation}
In the $\mathrm{SMOM}_{\gamma_{\mu}}$ renormalization scheme, the momentum is chosen as the symmetry form \cite{He:2022lse,Bi:2023pnf} to compute the projected vertex functions $\Gamma_{V}$:
\begin{equation}
    \begin{aligned}
    ap_{1} & =\dfrac{2\pi}{L}(n,n,0,0),\\
    ap_{2} & =\dfrac{2\pi}{L}(n,0,n,0),
\end{aligned}
\end{equation}
 where $n=1,2,3...$. The momentum transfer at the projected vertex is defined as $p=p_{1}-p_{2}$, satisfying the $\mathrm{SMOM}_{\gamma_{\mu}}$ scheme condition $p_{1}^{2}=p_{2}^{2}=p^{2}=\mu^{2}$. Since the projected vertex functions $\Gamma_{V}$ and $\Gamma_{O_{i}}$ are not gauge invariant, gauge fixing must be applied, and we adopt the Landau gauge.

 \subsection{Projected vertex function of baryon Operators}

 As a composite operator, the charmed baryon interpolating operator requires renormalization. Similar to the renormalization procedure for bilinear operators, the renormalization constant for the charmed baryon operator is defined as:
\begin{equation}
    \Gamma_{O_{i}}(p_{3},p_{4},p_{5})=P_{i}\,\Lambda_{O_{i}}(p_{3},p_{4},p_{5}),
    \label{eq: Gamma_Oi}
\end{equation}
where the definition of amputated Green’s function $\Lambda_{O_{i}}(p_{3},p_{4},p_{5})$ is:
\begin{equation}
    \begin{aligned}
    \Lambda_{O_{i}}(p_{3},p_{4},p_{5}) = & G_{O_{i}}(p_{3},p_{4},p_{5})\\
    \times &S^{-1}(p_{3})S^{-1}(p_{4})S^{-1}(p_{5}).
    \end{aligned}
\end{equation}
In the above equation, $G_{O_{i}}(p_{3},p_{4},p_{5})$ are the four-point Green's functions in momentum space:
 \begin{equation}
 \begin{aligned}
     G_{O_{i}}(p_{3},p_{4},p_{5})=&\sum_{x_{1}x_{2}x_{3}}e^{-i(p_{3}\cdot x_{1}+p_{4}\cdot x_{2}+p_{5}\cdot x_{3})}\\
     \times & \langle O_{i}(0),\,\bar{q}_{1}(x_{1})\bar{q}_{2}(x_{2})\bar{q}_{3}(x_{3})\rangle.
     \end{aligned}
 \end{equation}
Following ref.~\cite{Aoki:2006ib, Yoo:2021gql}, this work adopts a normalized projection after contracting the spinor and color indices of the charmed baryon operator, maximizing the overlap with the original operator. The projections for the charmed baryon operators $O_{1}\sim O_{3}$ are defined as follows:
\begin{equation}
    \begin{aligned}
    (P_{1})_{\alpha\beta\gamma\delta}^{ijk} & =\dfrac{1}{96}\epsilon_{ijk}(\gamma_{5}C)_{\alpha\beta}(\mathbbm{1})_{\gamma\delta},\\
    (P_{2})_{\alpha\beta\gamma\delta}^{ijk} & =\dfrac{1}{384}\epsilon_{ijk}(\gamma_{\mu}C)_{\alpha\beta}(\gamma_{5}\gamma^{\mu})_{\gamma\delta},\\
    (P_{3})_{\alpha\beta\gamma\delta}^{ijk} & =\dfrac{1}{384\sqrt{2}}\epsilon_{ijk}(\gamma_{\mu}C)_{\alpha\beta}(\gamma_{5}\gamma^{\mu})_{\gamma\delta},
\end{aligned}
\end{equation}
where Greek letters $\alpha,\beta,\gamma, \delta$ represents the spinor indices. 
For the renormalization scheme of baryon operators, we use the SYM3q scheme \cite{Yoo:2021gql}, and the momentum of the baryon operators in the Lattice QCD calculation are:
\begin{equation}
    \begin{aligned}
    ap_{3} & =\dfrac{2\pi}{L}(n,n,0,0),\\
    ap_{4} & =\dfrac{2\pi}{L}(-n,0,n,0),\\
    ap_{5} & =\dfrac{2\pi}{L}(0,-n,-n,0),
\end{aligned}
\end{equation}
where $n=1,2,3...$. Under the $\mathrm{SYM3q}$ renormalization scheme, the projected vertex momentum of the charmed baryon operators satisfies $k=p_{3}+p_{4}+p_{5}=0$ and the following condition is imposed $p_{3}^{2}=p_{4}^{2}=p_{5}^{2}=\mu^{2}$.

\subsection{Renormalization constants}
\begin{figure*}
\centering
\includegraphics[scale=0.38]{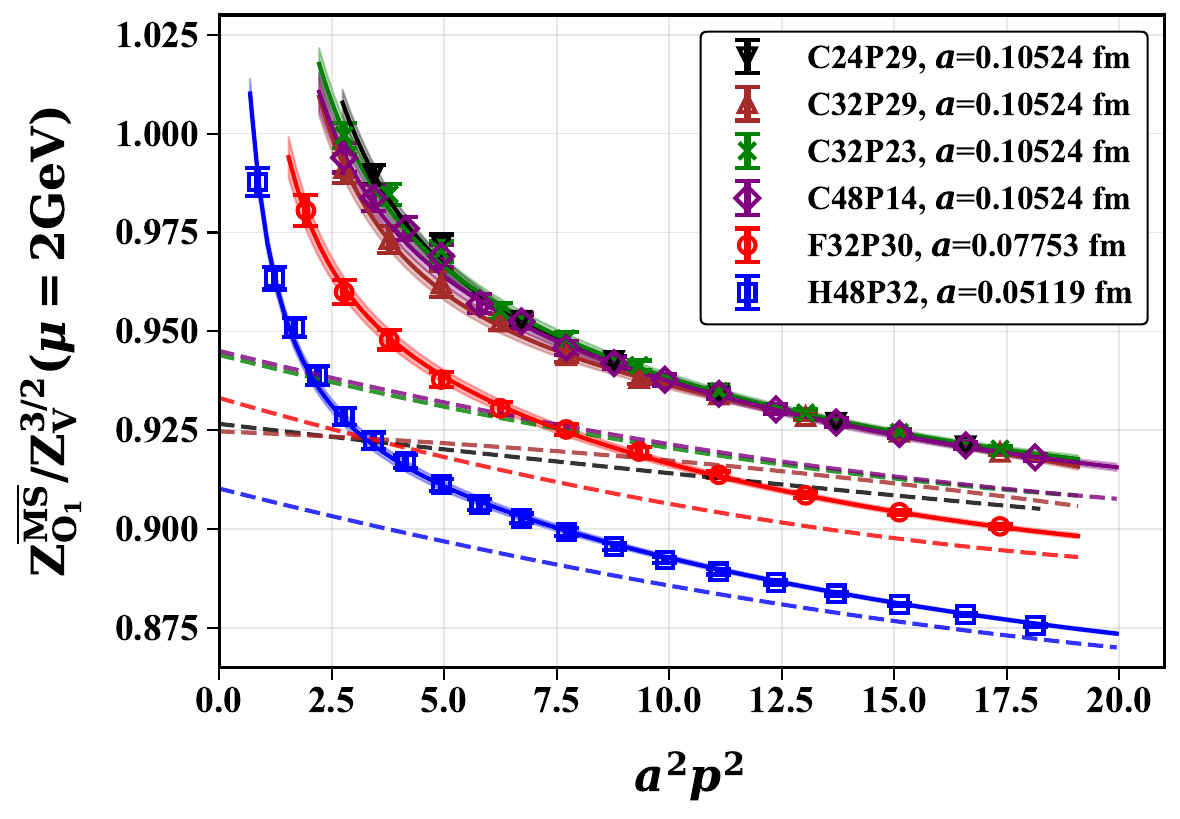}\qquad\qquad\quad
\includegraphics[scale=0.38]{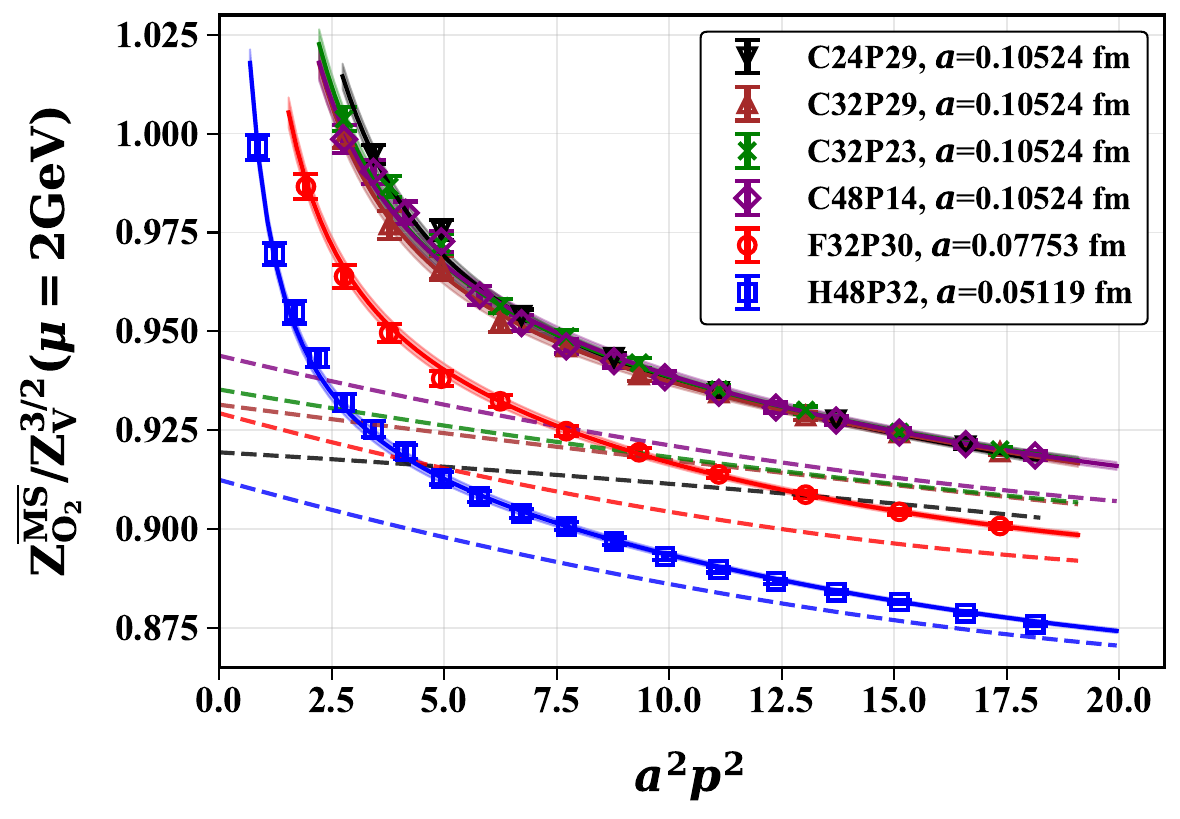}
\caption{The left and right panels show the results for the ratios of renormalization constants $Z_{O_{1}}^{\overline{\mathrm{MS}}}/Z_{V}^{3/2}$ and $Z_{O_{2}}^{\overline{\mathrm{MS}}}/Z_{V}^{3/2}$ at the scale $\mu = 2\mathrm{GeV}$, where the black, brown, green, purple, red, and blue error bars correspond to the C24P29, C32P29, C32P23, C48P14, F32P30, and H48P32 ensembles, respectively. {The solid lines show the fitting results of Eq.~(\ref{eq: a2p2 fit}). The dashed lines show the extrapolation results for $a^2p^2=0$ after removing the $c^{-1}(a^{2}p^{2})^{-1}$ term in Eq.~(\ref{eq: a2p2 fit}).}
}
\label{fig: a2p2}
\end{figure*}
 Considering the renormalization conditions of charmed baryon operators in Eq.~(\ref{eq: charmed baryon renormalization condition}) and bilinear operators in Eq.~(\ref{eq: bilinear renormalization condition}), we can construct the ratio of the renormalization constants as follows to avoid directly dealing with $Z_q$,
\begin{equation}
    \dfrac{Z^{\mathrm{SYM_{3q}}}_{O_{i}}}{(Z^{\mathrm{SMOM}_{\gamma_{\mu}}}_{V})^{3/2}}=\dfrac{\Gamma_{V}^{3/2}}{\Gamma_{O_{i}}}\bigg|_{\mathrm{SYM3q/SMOM_{\gamma_{\mu}}} }.
\end{equation}
The ratio $Z_{O_{i}}/Z_{V}^{3/2}$ is computed at the finite quark mass on the lattice. Aiming at a mass-independent renormalization scheme, the chiral extrapolation is required and we use a linear extrapolation form \cite{He:2022lse, Bi:2023pnf, CLQCD:2023sdb}: 
\begin{equation}
    {\dfrac{Z^{\mathrm{SYM_{3q}}}_{O_{i}}}{(Z^{\mathrm{SMOM_{\gamma_{\mu}}}}_{V})^{3/2}}(m_{\pi}^{2})=B+Cm^2_{\pi},}
    \label{eq: mpi^2 extrapolation}
\end{equation}
where the coefficient $B$ represents the extrapolation result in the chiral limit. After performing the chiral extrapolation, a conversion factor is required to transform the ratio of the renormalization constants from the $\mathrm{SMOM}_{\gamma_{\mu}} / \mathrm{SYM3q}$ scheme to the $\overline{\mathrm{MS}}$ scheme:
\begin{equation}
 \begin{aligned}
     \dfrac{Z^{\mathrm{\overline{MS}}}_{O_{i}}}{Z_{V}^{3/2}}(a^{2}p^{2},\mu)=&C^{\mathrm{\overline{\mathrm{MS}}\leftarrow\mathrm{SMOM}_{\gamma_{\mu}}/SYM3q}}(\mu)\\
     \times&\dfrac{Z^{\mathrm{SYM3q}}_{O_{i}}}{(Z^{\mathrm{SMOM}}_{V})^{3/2}}(a^{2}p^{2}).
\end{aligned}
\label{eq: a2p2 extrapolation}
\end{equation}
In the $\mathrm{SMOM_{\gamma_{\mu}}}$ scheme, the vector Ward-Takahashi identity guarantees $Z_{V}=1$ in the continuum limit~\cite{Sturm:2009kb}, so the conversion factor $C^{\mathrm{\overline{MS}}}_{V}=1$. The conversion factor of charmed baryon operators in Eq.~(\ref{eq:O1-O3}) are given by \cite{Yoo:2021gql}:
\begin{equation}
    \begin{aligned}
C^{\mathrm{\overline{\mathrm{MS}}\leftarrow\mathrm{SMOM}_{\gamma_{\mu}}/SYM3q}}=\Lambda^{\overline{\mathrm{MS}}}_{\mathrm{SYM3q}} \big(C_{q}^{\overline{\mathrm{MS}}\leftarrow\mathrm{SMOM}_{\gamma_{\mu}}}\big)^{3/2},
    \end{aligned}
    \label{eq: conversion factor of O1}
\end{equation}
where the quark field renormalization constant to the next-to-leading order (NLO) are given in Ref.~\cite{Almeida:2010ns}:
\begin{equation}
C_{q}^{\overline{\mathrm{MS}}\leftarrow\mathrm{SMOM}_{\gamma_{\mu}}}=1+C_{F}\dfrac{\alpha_{s}}{4\pi}+O(\alpha_{s}^{2}).
\end{equation}
The conversion factor $ C^{\overline{\mathrm{MS}} \leftarrow \mathrm{SMOM}_{\gamma_{\mu}} / \mathrm{SYM3q}}(\mu) $ serves to convert the renormalization constant $Z_{O_i}$ of the charmed baryon operator from the $\mathrm{SMOM_{\gamma_{\mu}}/SYM3q}$ scheme to the $\overline{\mathrm{MS}}$ scheme.
For the pseudoscalar-current operator $O_{1}$, this work incorporates next-to-leading order corrections, ensuring the absence of mixing among operators. The explicit expression for the conversion factor is provided in Ref.~\cite{Gracey:2012gx}:
 \begin{equation}
    \Lambda^{\overline{\mathrm{MS}}}_{\mathrm{SYM3q}}=1+0.989426\dfrac{\alpha_{s}}{4\pi}+O(\alpha_{s}^{2}).
     \label{eq: amputated green function in NLO}
 \end{equation}
 At the one-loop level, the perturbative calculation results of baryon operators without mixing \cite{Gracey:2012gx}. The anomalous dimension of operator $O_{1}$ is provided in Ref.~\cite{Gracey:2012gx}:
 \begin{equation}
     \gamma=-\dfrac{\alpha_{s}}{2\pi}+O(\alpha_{s}^{2}).
     \label{eq: anomalous dimension}
 \end{equation}
For the vector-current operators $O_{2}$ and $O_{3}$, we use the Fierz transformation formula provided in Ref.~\cite{QCDSF:2008qtn} to obtain:
\begin{equation}
    O_{2} =2\gamma_{5}(O_{a}+O_{b}),
\end{equation}
where
\begin{equation}
    \begin{aligned}
    O_{a} & =\epsilon_{ijk}(q_{1,i}^{T}Cq_{2,j})q_{3,k},\\
    O_{b} & =\epsilon_{ijk}(q_{1,i}^{T}C\gamma_{5}q_{2,j})\gamma_{5}q_{3,k}.
\end{aligned}
\end{equation}
\begin{table}
    \renewcommand{\arraystretch}{1.8}
  \setlength{\tabcolsep}{1.3mm}
\centering
\caption{Renormalization constants of bilinear operator at different ensembles in Ref.~\cite{CLQCD:2024yyn}}
\begin{tabular}{ccccc}
\toprule 
Symbol & $a$ (fm) & $Z_{V}^{l}$ & $Z_{V}^{s}$ & $Z_{V}^{c}$\tabularnewline
\hline
C24P29 & 0.10524 & 0.79814(23) & 0.85184(06) & 1.57353(18)\tabularnewline
C32P29 & 0.10524 & 0.79810(13) & 0.85167(04) & 1.57163(14)\tabularnewline
C32P23 & 0.10524 & 0.79957(13) & 0.85350(04) & 1.57644(12)\tabularnewline
C48P14 & 0.10524 & 0.79957(06) & 0.85359(02) & 1.57415(08)\tabularnewline
F32P30 & 0.07753 & 0.83548(12) & 0.86900(03) & 1.30566(07)\tabularnewline
H48P32 & 0.05199 & 0.86855(04) & 0.88780(01) & 1.12882(11)\tabularnewline
\toprule 
\end{tabular}
\label{tab: ZV}
\end{table}

In the above equations, the amputated green function $\Lambda_{\mathrm{SYM3q}}^{\mathrm{\overline{MS}}}$ of $O_{a}$ and $O_{b}$ in one-loop result are given in $\mathrm{\overline{MS}}$ \cite{Gracey:2012gx}, which are the same as Eq.~(\ref{eq: amputated green function in NLO}), since the effect of $\gamma_{5}$ appears will not appear at NLO \cite{Gracey:2012gx}. Therefore the conversion factors of operator $O_{2}$ and $O_{3}$ are the same as $O_{1}$ in Eq.~(\ref{eq: conversion factor of O1}), and the anomalous dimensions of $O_{2}$ and $O_{3}$ are the same as $O_{1}$ in Eq.~(\ref{eq: anomalous dimension}). 
After evolving the ratio of renormalization constants in Eq.~(\ref{eq: a2p2 extrapolation}) to the scale $\mu=2~\mathrm{GeV}$, we perform an extrapolation in $a^{2}p^{2}$ for $Z^{\mathrm{\overline{MS}}}_{O_{i}}/Z_{V}^{3/2}(2\mathrm{GeV},a^{2}p^{2})$ using the formula \cite{Bi:2023pnf, He:2022lse}:
\begin{equation}
    \begin{aligned}
        \dfrac{Z_{O_{i}}^{\mathrm{\overline{MS}}}}{Z_{V}^{3/2}}(a^{2}p^{2},2\mathrm{GeV})=&\dfrac{Z^{\mathrm{\overline{MS}}}_{O_{i}}}{Z_{V}^{3/2}}(\mathrm{2GeV})+c_{-1}(a^{2}p^{2})^{-1}\\
        +&c_{1}(a^{2}p^{2})+c_{2}(a^{2}p^{2})^{2}.
    \end{aligned}
    \label{eq: a2p2 fit}
\end{equation}
The extrapolated results for $a^{2} p^{2}$ are shown in Fig.~\ref{fig: a2p2} The left panel corresponds to the extrapolated results for $Z_{O_{1}}^{\overline{\mathrm{MS}}} / Z_{V}^{3/2}$, while the right panel shows the extrapolated results for $Z_{O_{2}}^{\overline{\mathrm{MS}}} / Z_{V}^{3/2}$.  {The solid lines show the fitting results of Eq.~(\ref{eq: a2p2 fit}) and the dashed lines show the extrapolation results after removing the $c^{-1}(a^{2}p^{2})^{-1}$ term in Eq.~(\ref{eq: a2p2 fit}).} The extrapolated results for the operator $O_{3}$ are found to be consistent with those of $O_{2}$ in the chiral limit. {Since the chiral extrapolation of valence quark masses is performed for the renormalization constants on each ensemble, the renormalization constants for the ensembles with the same lattice spacing are close as shown in Fig.~\ref{fig: a2p2}. The minor differences are mainly due to the distinct sea quark masses of different ensembles.}

After performing the $a^{2}p^{2}$ extrapolation, we multiply the ratio $Z^{\mathrm{\overline{MS}}}_{O_{i}}/Z_{V}^{3/2}(\mathrm{2GeV})$ by the renormalization constant of the vector current operator:
\begin{equation}
    Z^{\mathrm{\overline{MS}}}_{\mathcal{B}_{c}}(\mathrm{2GeV})=\dfrac{Z^{\mathrm{\overline{MS}}}_{O_{i}}}{Z_{V}^{3/2}}(\mathrm{2GeV})\big(Z_{V}^{q_{1}}Z_{V}^{q_{2}}Z_{V}^{q_{3}}\big)^{1/2},
\end{equation}
where the numerical results of $Z_{V}^{l},Z_{V}^{s},Z_{V}^{c}$ at different ensembles are given in Ref.~\cite{CLQCD:2024yyn}, which are shown in Tab.~\ref{tab: ZV}
The numerical values of the renormalization constants $Z^{\mathrm{\overline{MS}}}_{\mathcal{B}_{c}}(a,\mu)$ for the charmed baryon decay constants are provided in Tab.~\ref{tab: ZO}.

{To estimate the systematic uncertainty from the renormalization constants, we adopt the following procedures: for the chiral extrapolation in Eq.~(\ref{eq: mpi^2 extrapolation}), we exclude the lightest mass point to assess the uncertainty; for $a^{2}p^{2}$ extrapolation in Eq.~(\ref{eq: a2p2 fit}), we remove the data corresponding to the minimum value of \(a^{2}p^{2}\) as one of the systematic error, the another systematic error is estimated by drop the $c_{-1}(a^{2}p^{2})^{-1}$ term in Eq.~(\ref{eq: a2p2 fit}). Additionally, the systematic uncertainty associated with the conversion factor in Eq.~\ref{eq: conversion factor of O1} is estimated by the two-loop results in Refs.~\cite{Almeida:2010ns, Gracey:2012gx, Yoo:2021gql}.}

\begin{figure*}
\centering
\includegraphics[scale=0.28]{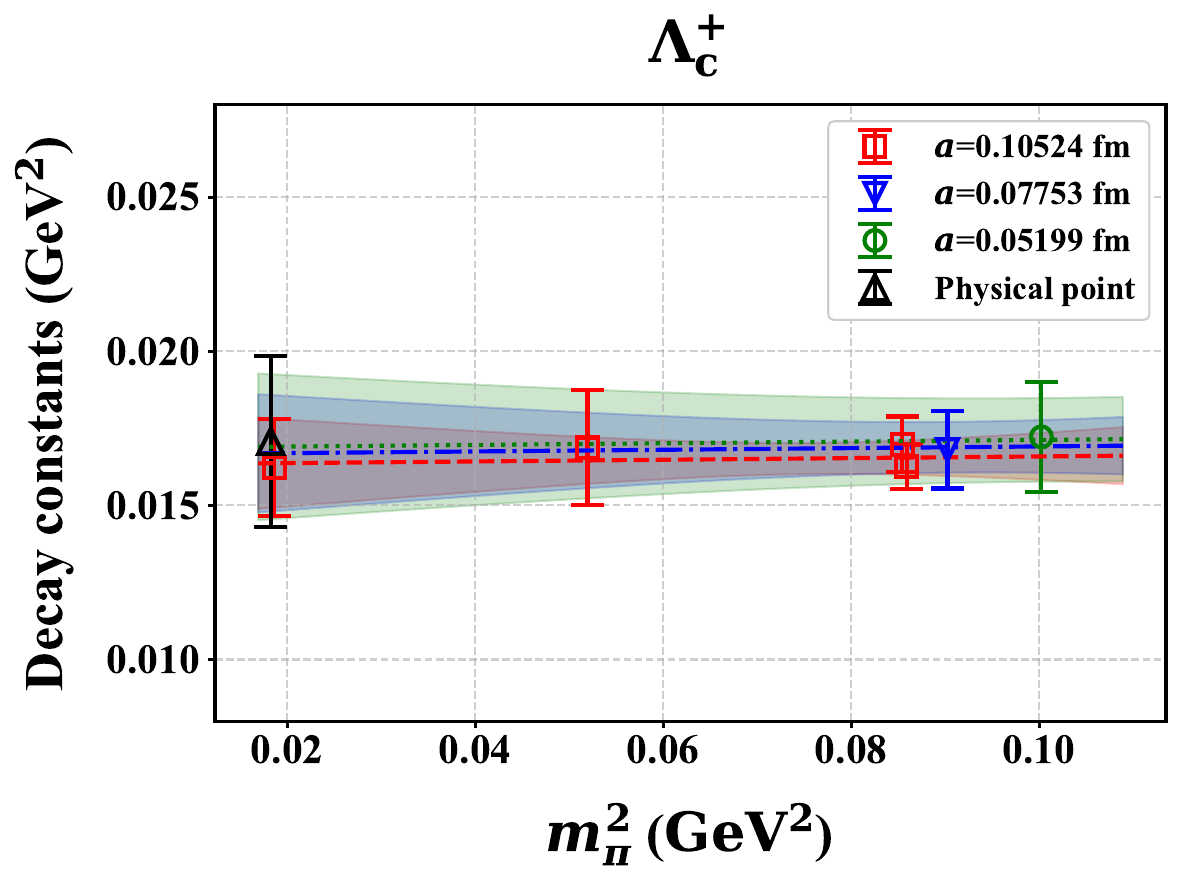}$\quad$
\includegraphics[scale=0.28]{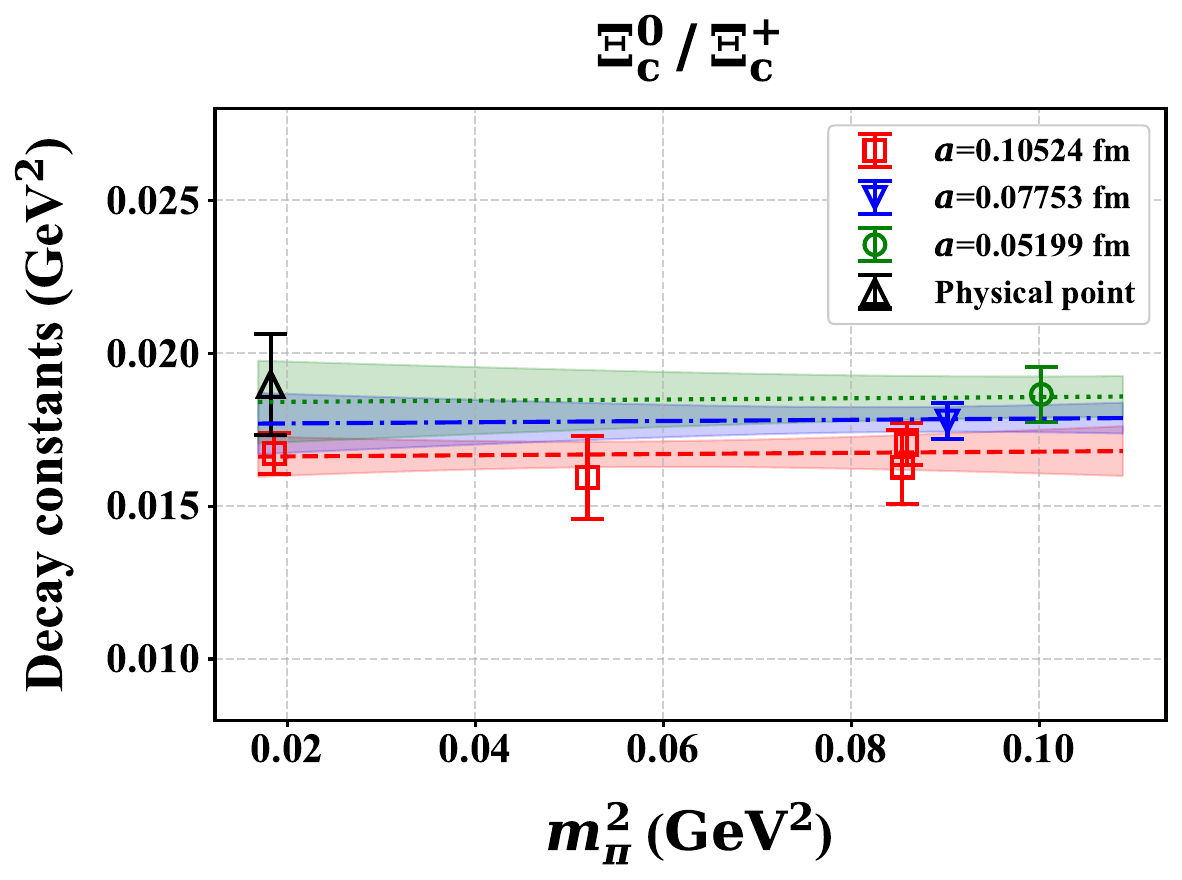}$\quad$
\includegraphics[scale=0.28]{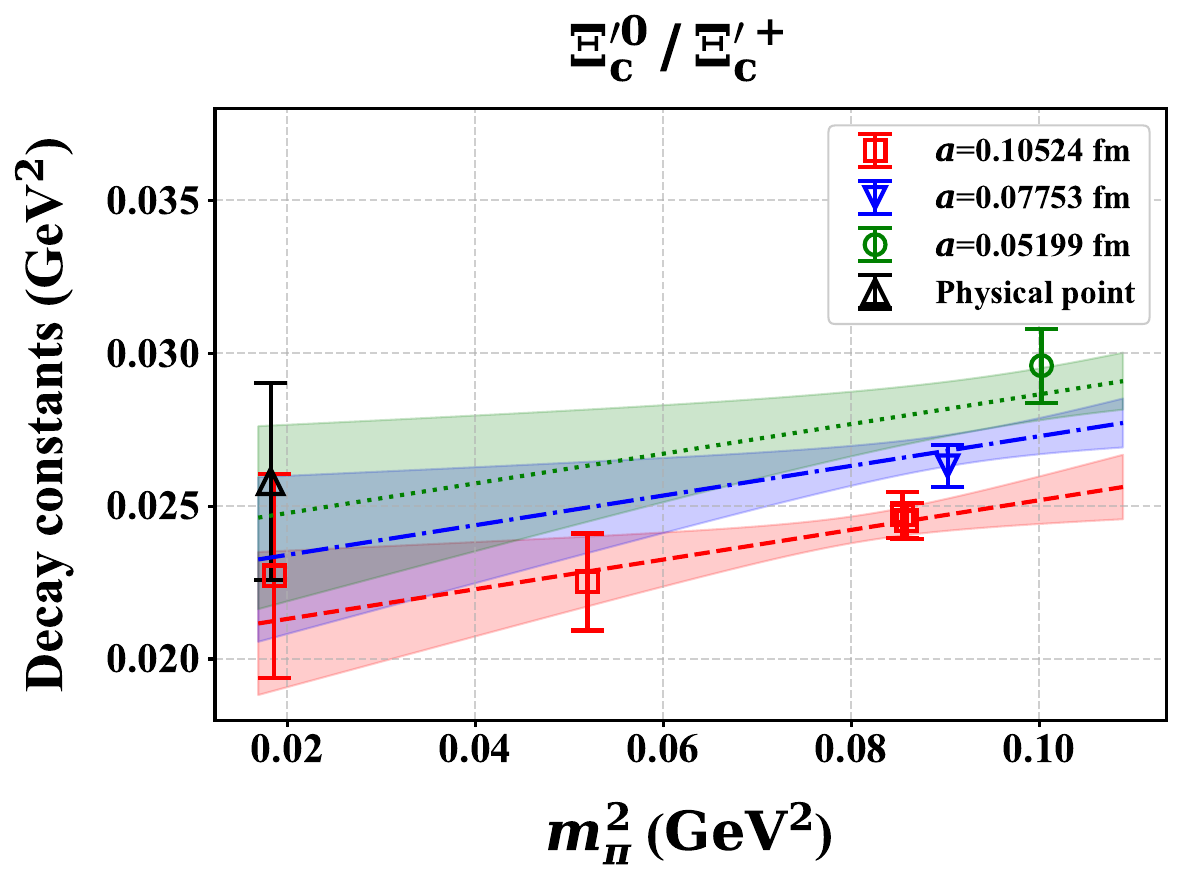}\\
\includegraphics[scale=0.28]{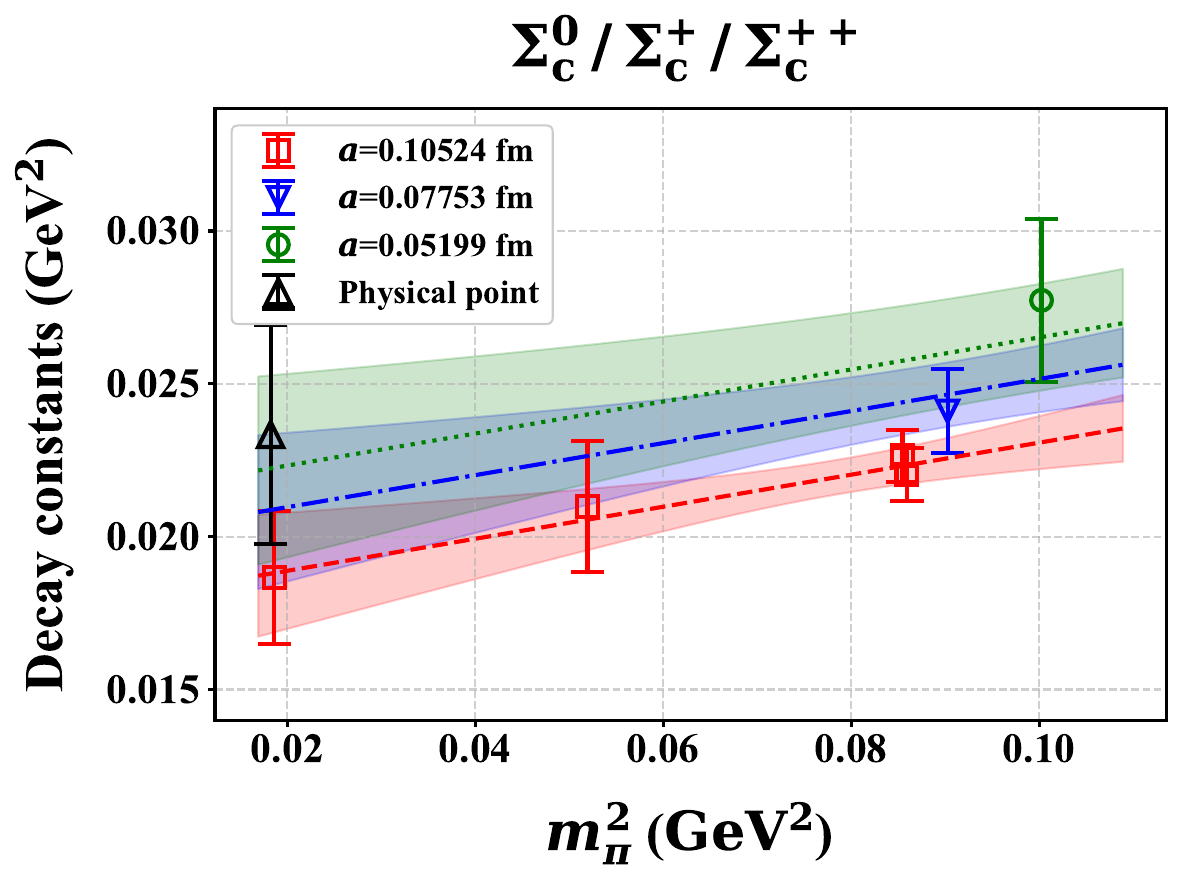}$\quad$
\includegraphics[scale=0.28]{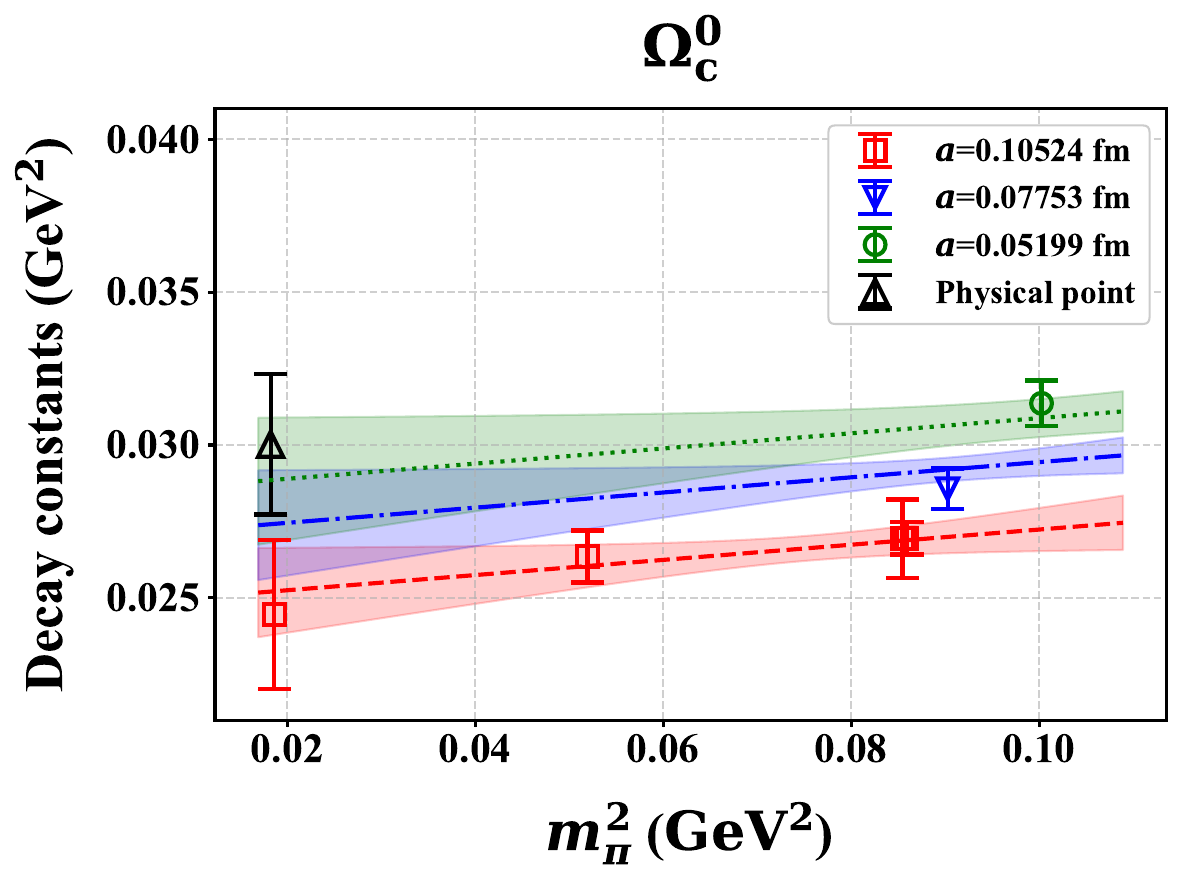}
\caption{The chiral and continuum extrapolation results for the charmed baryon decay constants are shown as a function of $m_{\pi}^{2}$. The red, green, and blue error bands correspond to lattice spacings of 0.10524 fm, 0.07753 fm, and 0.05199 fm, respectively, while the black error bar represents the physical point.
}
\label{fig: extrapolation}
\end{figure*}
\begin{table*}
    \renewcommand{\arraystretch}{2.0}
  \setlength{\tabcolsep}{2.5mm}
\centering
\caption{Renormalization constants of charmed baryons at $\mathrm{\mu=2GeV}$ for different ensemble.}
\begin{tabular}{cccccc}
\toprule 
Ensemble & $\quad Z^{\overline{\mathrm{MS}}}_{\Lambda_{c}}\quad$ & $\quad Z^{\overline{\mathrm{MS}}}_{\Xi_{c}}\quad$ & $\quad Z^{\overline{\mathrm{MS}}}_{\Xi_{c}^{\prime}}\quad$ & $\quad Z^{\overline{\mathrm{MS}}}_{\Sigma_{c}}\quad$ & $\quad Z^{\overline{\mathrm{MS}}}_{\Omega_{c}}\quad$\tabularnewline
\hline
C24P29 & 0.9277(82) & 0.9584(84) & 0.9510(85) & 0.9205(82) & 0.9824(88) \tabularnewline
C32P29 & 0.9252(91) & 0.9557(95) & 0.963(10) & 0.9319(98) & 0.994(10) \tabularnewline
C32P23 & 0.9477(85) & 0.9792(88) & 0.9701(84) & 0.9389(81) & 1.0023(86) \tabularnewline
C48P14 & 0.9481(86) & 0.9796(89) & 0.9783(80) & 0.9469(77) & 1.0108(83) \tabularnewline
F32P30 & 0.8908(56) & 0.9085(57) & 0.9048(51) & 0.8872(50) & 0.9228(52)\tabularnewline
H48P32 & 0.8400(27) & 0.8492(28) & 0.8513(26) & 0.8420(26) & 0.8607(27)\tabularnewline
\toprule 
\end{tabular}
\label{tab: ZO}
\end{table*}

\section{Numerical Results}
\label{sec:4}

\begin{figure}
\centering
		\scalebox{0.45}[0.45]{\includegraphics[width=1\textwidth]{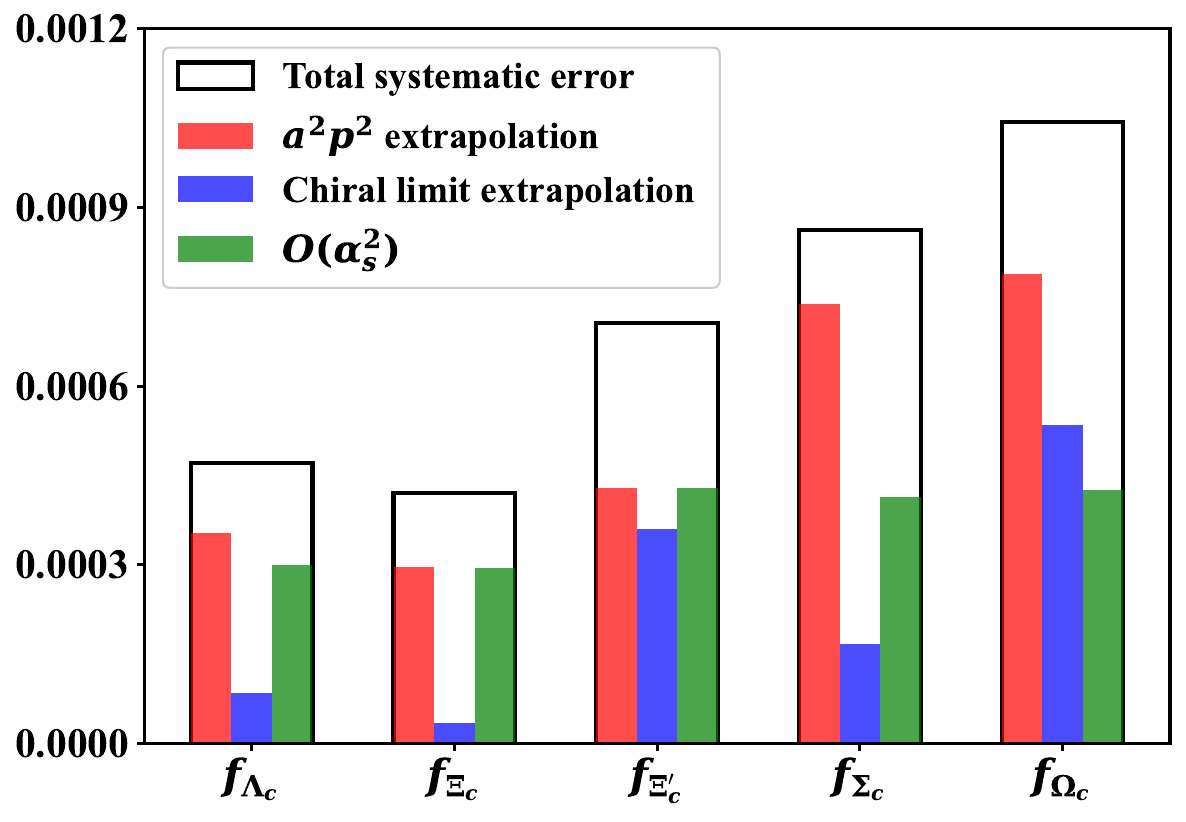}}
		\caption{{Sources of systematic uncertainties for the decay constant of charmed baryons. The red, blue, green, and white bars denote the systematic errors from the $a^{2}p^{2}$ extrapolation, the chiral extrapolation, the $O(\alpha_{s}^{2})$ conversion factor, and the total systematic uncertainty, respectively. }}
\label{fig: systematic error}
\end{figure}

\begin{figure*}
\centering
\includegraphics[scale=0.75]{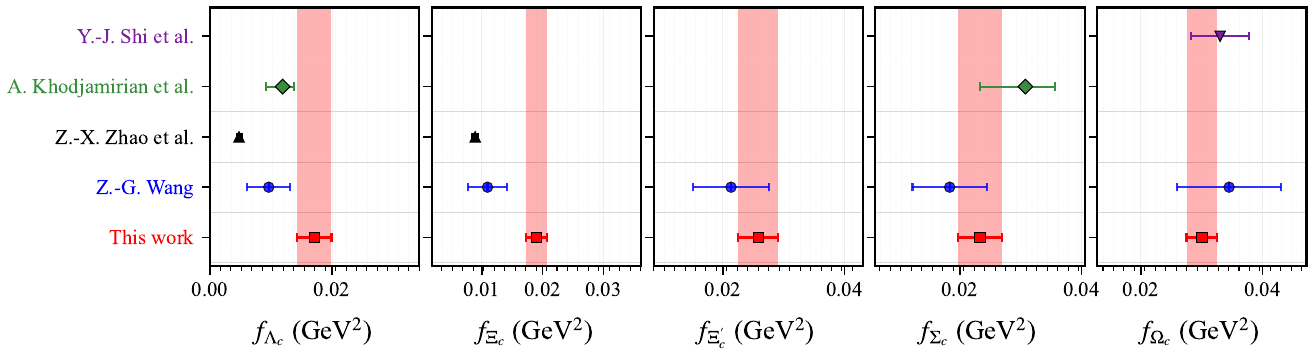}
\caption{Comparison of the numerical results for the charmed baryon decay constants $\Lambda_{c}, \Xi_{c}, \Xi_{c}^{\prime}, \Sigma_{c}, \Omega_{c}$. The red points represent the results from the lattice QCD calculations in this work, while the other error bars correspond to QCD sum rule results \cite{Wang:2009cr, Wang:2010fq, Shi:2019hbf, Zhao:2020mod, Khodjamirian:2011jp}.}

\label{fig: compare with sum rule}
\end{figure*}

After renormalizing the bare decay constants $f_{\mathcal{B}_{c}}^{(0)}$ using the renormalization constants $Z^{\mathrm{\overline{MS}}}_{\mathcal{B}_{c}}$ at $\mu = 2$GeV provided in Tab.~\ref{tab: ZO}, we obtain the renormalized decay constants $f_{\mathcal{B}_{c}}$ in different ensembles. To obtain the decay constants at the physical mass point and in the continuum limit, we perform lattice spacing and chiral extrapolations using the following formula:
\begin{equation}
    f_{\mathcal{B}_{c}}=f_{\mathcal{B}_{c},\mathrm{phy}}+c_{1}\left(m_{\pi}^{2}-m_{\pi,phy}^{2}\right)+c_{2}a^{2},
\end{equation}
where $f_{\mathcal{B}_{c}}$ denotes the decay constant on each ensemble, while $f_{\mathcal{B}_{c},\mathrm{phy}}$ represents the decay constant after extrapolation. The extrapolation results are shown in the Fig.~\ref{fig: extrapolation}. The chiral and continuum extrapolation results for the charmed baryon decay constants are presented as functions of the $m_{\pi}^{2}$, with the red, green, and blue error bands indicating lattice spacings of 0.10524 fm, 0.07753 fm, and 0.05199 fm, respectively, and the black error bar denoting the extrapolated value at the physical point. The numerical results of charmed baryon decay constants at $\mu=2\mathrm{GeV}$ are given in follow:
\begin{equation}
\begin{aligned}
    f_{\Lambda_{c}} =& 1.71(28)_{\text{stat.}}(5)_{\text{syst.}} \times 10^{-2} \,\mathrm{GeV}^{2},\\
    f_{\Xi_{c}} =& 1.90(16)_{\text{stat.}}(4)_{\text{syst.}} \times 10^{-2} \,\mathrm{GeV}^{2},\\
    f_{\Xi^{\prime}_{c}} =& 2.58(32)_{\text{stat.}}(7)_{\text{syst.}} \times 10^{-2} \,\mathrm{GeV}^{2},\\
    f_{\Sigma_{c}} =& 2.33(36)_{\text{stat.}}(9)_{\text{syst.}} \times 10^{-2} \,\mathrm{GeV}^{2},\\
    f_{\Omega_{c}} =& 3.00(23)_{\text{stat.}}(10)_{\text{syst.}} \times 10^{-2} \,\mathrm{GeV}^{2},
    \label{eq: decay constants}
\end{aligned} 
 \end{equation}
{where the systematic uncertainties come from the nonperturbative renormalization as discussed in Sec.\ref{sec:3}. Fig.~\ref{fig: systematic error} illustrates the various sources of systematic uncertainty in the decay constant of charmed baryons. The red, blue, green, and white bars correspond to systematic uncertainties arising from the $a^{2}p^{2}$ extrapolation, the chiral extrapolation, the $O(\alpha_{s}^{2})$ conversion factor, and the total systematic error, respectively.}

The $\chi^{2}/\mathrm{d.o.f.}$ values for the above decay constants range from 0.08 to 0.26, and the results indicate that the precision of the charmed baryon decay constants lies within $8\sim16\%$. Fig.~\ref{fig: compare with sum rule}  presents a comparison between the results from lattice QCD and those from QCD sum rules with the same interpolating current. {The relationship for the definition of decay constant between Ref.~\cite{Wang:2009cr, Wang:2010fq,  Shi:2019hbf, Zhao:2020mod} and this work is given by
$\lambda_{\mathcal{B}_{c}} = m_{\mathcal{B}_{c}} f_{\mathcal{B}_{c}}$.
For a direct comparison, we divide $\lambda_{\mathcal{B}_{c}}$ by the charmed baryon mass $m_{\mathcal{B}_{c}}$ as provided by the PDG \cite{ParticleDataGroup:2024cfk}.}

\section{Conclusion}
\label{sec:5}

In this work, we compute the decay constants of charmed baryons in lattice QCD. Using the di-quark SU(3) symmetry, we construct the operators for the charmed baryon anti-triplet and sextet. The two-point correlation functions of these operators are calculated on CLQCD gauge configurations, and the bare decay constants are extracted through two-state fits. Since the bare decay constants require further renormalization, we adopt the {SYM3q/$\mathrm{SMOM_{\gamma_{\mu}}}$} renormalization scheme to determine the charmed baryon renormalization constants at the scale $\mu = 2\mathrm{GeV}$ for different ensembles. After renormalization, lattice-spacing and pion-mass extrapolations are performed to obtain the charmed baryon decay constants with a precision of $8 \sim 16\%$, providing high-precision inputs from first principles for theoretical calculations.

\section*{Acknowledgments}

We thank Long Chen, Hai-Yang Du, Bo-Lun Hu, Mu-Hua Zhang for useful discussions. 
We thank the CLQCD collaborations for providing us the gauge configurations with dynamical fermions \cite{CLQCD:2023sdb}, which are generated on the HPC Cluster of ITP-CAS, the Southern Nuclear Science Computing Center (SNSC), the Siyuan-1 cluster supported by the Center for High Performance Computing at Shanghai Jiao Tong University and the Dongjiang Yuan Intelligent Computing Center.
This work is supported in part by Natural Science Foundation of China under grant No. 12375069, No. 12447183, No. 12475095, No. 12575084, No. 12447185, No.  12293060, No. 12525504, No. 12293062, No. 12435002, No. 12447101, No. 12447102 and No. 12175030. 
Q.A.Z is also supported by the Fundamental Research Funds for the Central Universities.
J.H is also supported by Guang-dong Major Project of Basic and Applied Basic Research No.2025A1515012199. This work is also supported in part by National Key R\&D Program of China No.2024YFE0109800.
The computations in this paper were run on the Siyuan-1 cluster supported by the Center for High Performance Computing at Shanghai Jiao Tong University, and Advanced Computing East China Sub-center.
The LQCD simulations were performed using the PyQUDA software suite \cite{Jiang:2024lto} and QUDA \cite{Clark:2009wm,Babich:2011np,Clark:2016rdz} through HIP programming model \cite{Bi:2020wpt}.

\section*{Appendix}
\label{Appendix}
In lattice QCD, the extraction of physical quantities relies on the selection of data points. To address systematic uncertainties from different model selections, we employ the Akaike Information Criterion (AIC)~\cite{Jay:2020jkz}, also referred to as model averaging. This method yields a probability-weighted combination of variations from different model choices, incorporating ``systematic errors'' due to model selection in a conservative manner.

In our implementation, we consider a set of models ${M}$ defined by varying the minimum time slice~$t_{\text{min}}\in [t_{\text{1,min}},t_{\text{2,min}}]$ included in the fit to the data~${D}$, while keeping~$t_{\text{max}}$ fixed. For each~$t_{\text{min}}$, we perform a fully correlated fit over the interval, generating an ensemble of model variations. The relative weight of each model is given by its posterior probability on the data \cite{Jay:2020jkz}:
\begin{equation}
\mathrm{pr}(M|D) \approx \exp\left[-\frac{1}{2}\left(\chi^{2}_{\mathrm{aug}}(\mathbf{a}^{\thinstar}) + 2k + 2N_{\mathrm{cut}}\right)\right],
\label{eq:model_weight}
\end{equation}
where $\chi^{2}_{\mathrm{aug}}(\mathbf{a}^{\thinstar})$ is the augmented chi-squared at the best-fit parameters $\mathbf{a}^{\thinstar}$, $k$ is the number of fit parameters, and $N_{\mathrm{cut}}$ is the number of data points excluded from the fit (i.e., with $t < t_{\mathrm{min}}$). The probabilities are normalized such that $\sum_M \mathrm{pr}(M|D) = 1$.

The model-averaged estimate for the mean of $a_{0}$ is \cite{Jay:2020jkz}
\begin{equation}
\langle a_{0} \rangle = \sum_{M} \langle a_{0} \rangle_M \, \mathrm{pr}(M|D),
\label{eq:model_average_mean}
\end{equation}
where $\langle a \rangle_M$ is the estimate from model~$M$. The corresponding uncertainty decomposes as \cite{Jay:2020jkz}
\begin{equation}
\begin{aligned}
\sigma_{a_{0}}^{2} = &\sum_{i=1}^{N_{M}}\sigma_{a_{0},i}^{2}\,\mathrm{pr}(M_{i}|D)+\sum_{i=1}^{N_{M}}\langle a_{0}\rangle_{i}^{2}\,\mathrm{pr}(M_{i}|D)\\
                   - &\bigg(\sum_{i=1}^{N_{M}}\langle a_{0}\rangle_{i} \,\mathrm{pr}(M_{i}|D) \bigg)^{2}
\label{eq:model_average_variance}
\end{aligned}
\end{equation}
The first term is the weighted average of the statistical variances, and the remaining terms capture the ``systematic errors'' from different model choice.

In this work, model averaging is applied to extract the ground-state masses of charmed baryons from two-point correlation functions. For each baryon and ensemble, we consider a range of fit windows $[t_{\text{min}}, t_{\text{max}}]$. 

We use a two-state fit ansatz incorporating both the ground state and the first excited state. For each $t_{\text{min}}$, a fully correlated fit is performed to extract effective mass $m_{\mathcal{B}_{c}}$ and the bare decay constants $f^{(0)}_{\mathcal{B}_{c}}$ according to Eq.~(\ref{eq: C2pt}) with different ensembles.

\begin{figure*}
\centering
\includegraphics[scale=0.25]{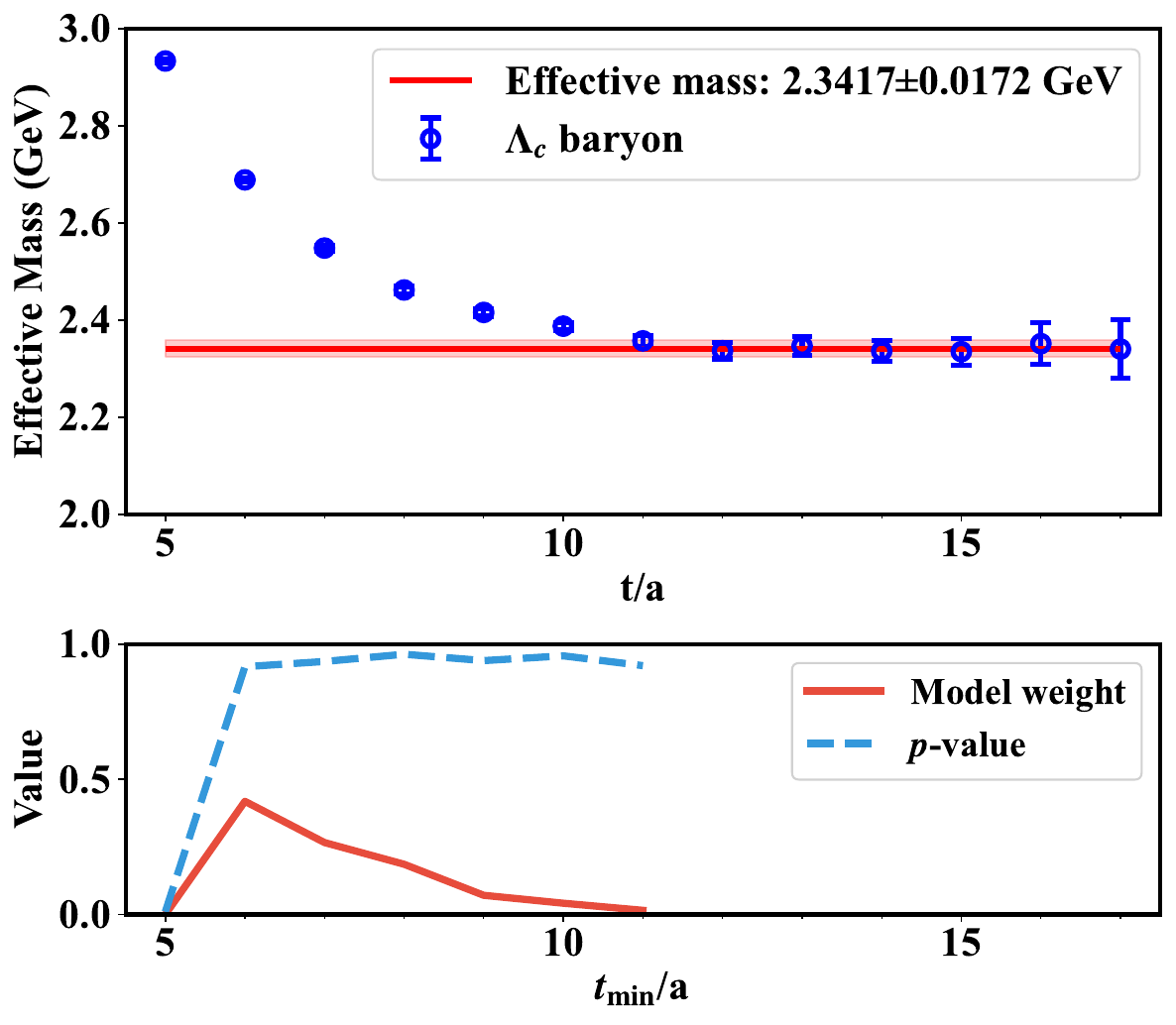}$\qquad$
\includegraphics[scale=0.25]{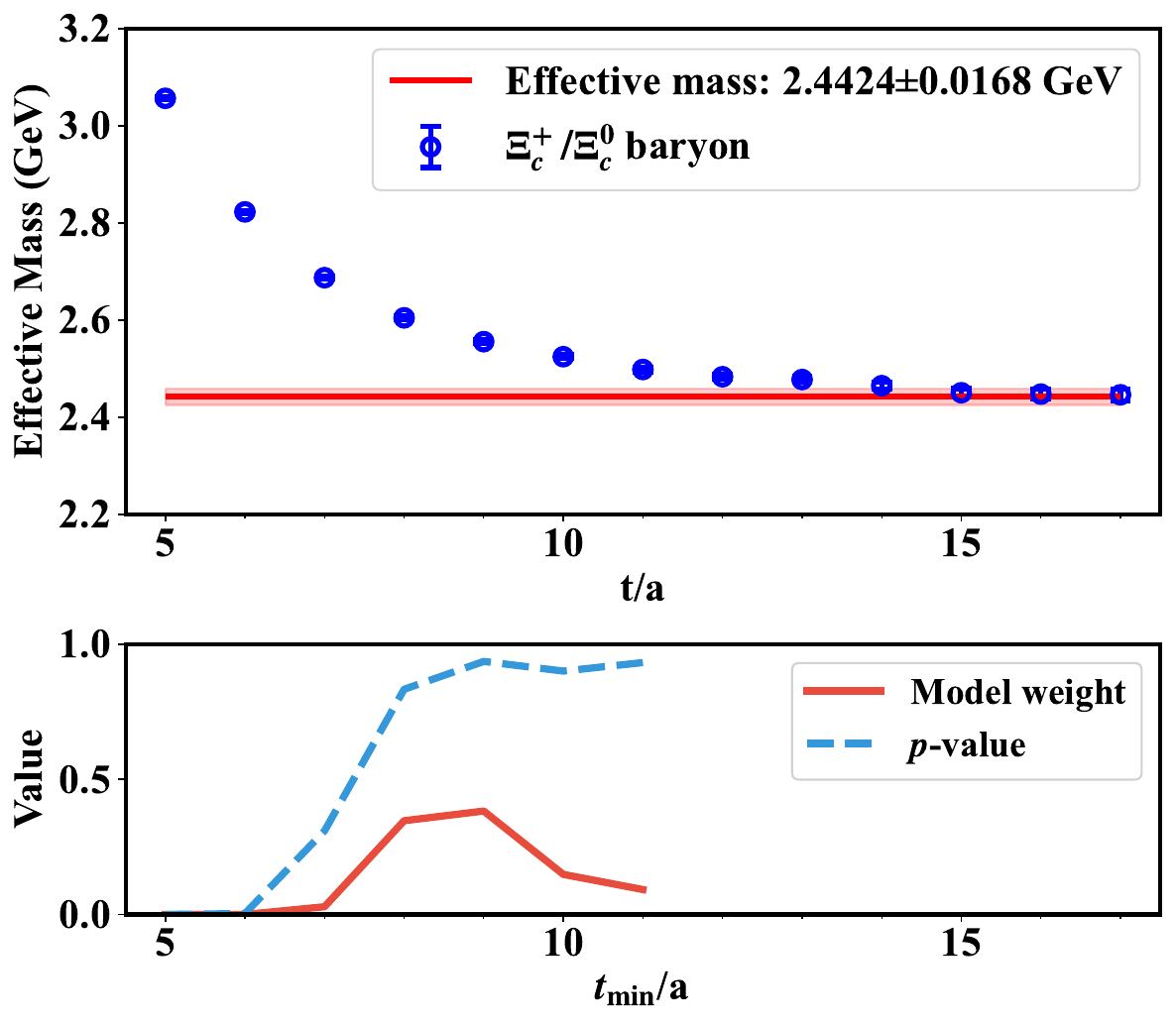}$\qquad$
\includegraphics[scale=0.25]{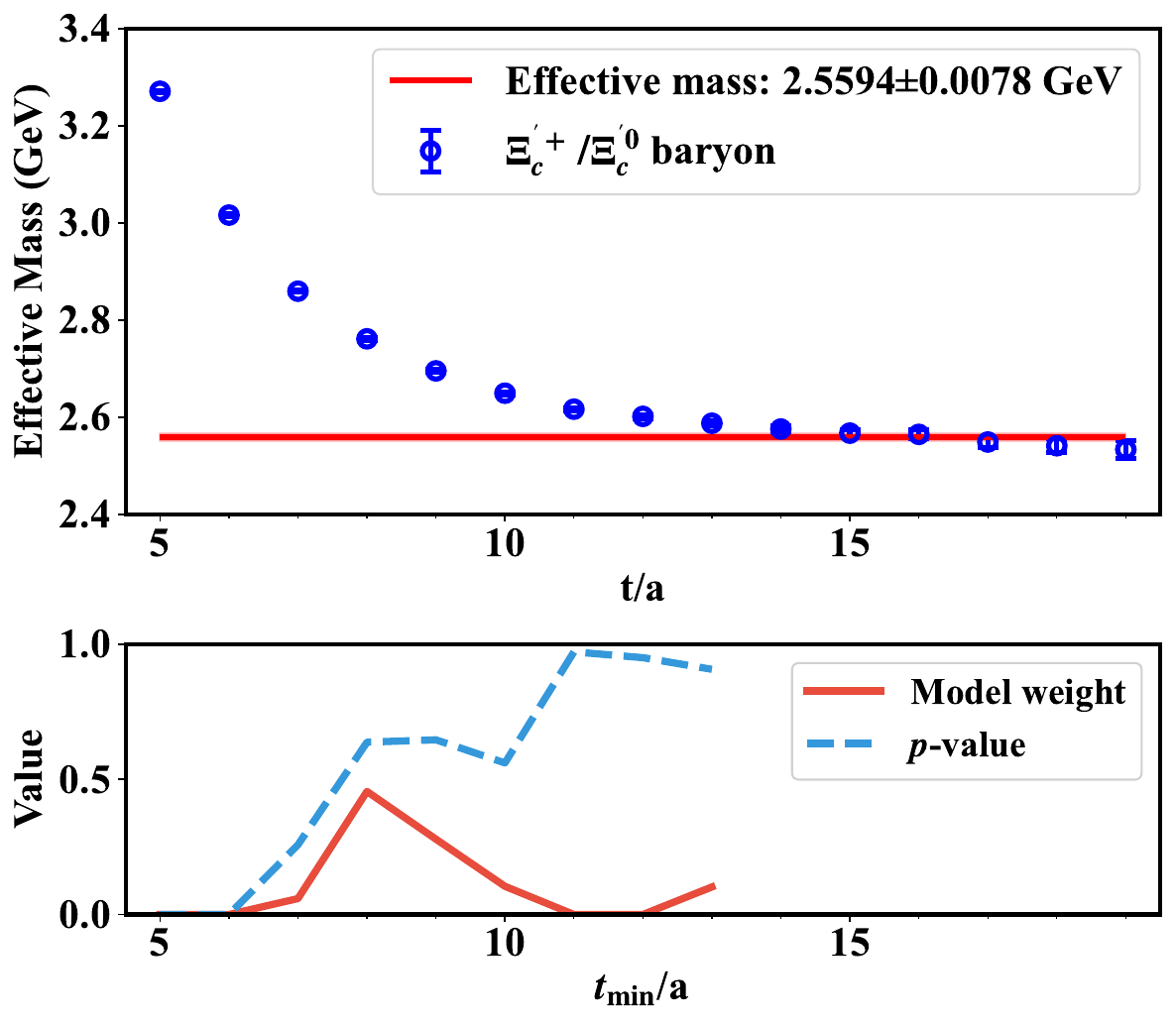}\\
\includegraphics[scale=0.25]{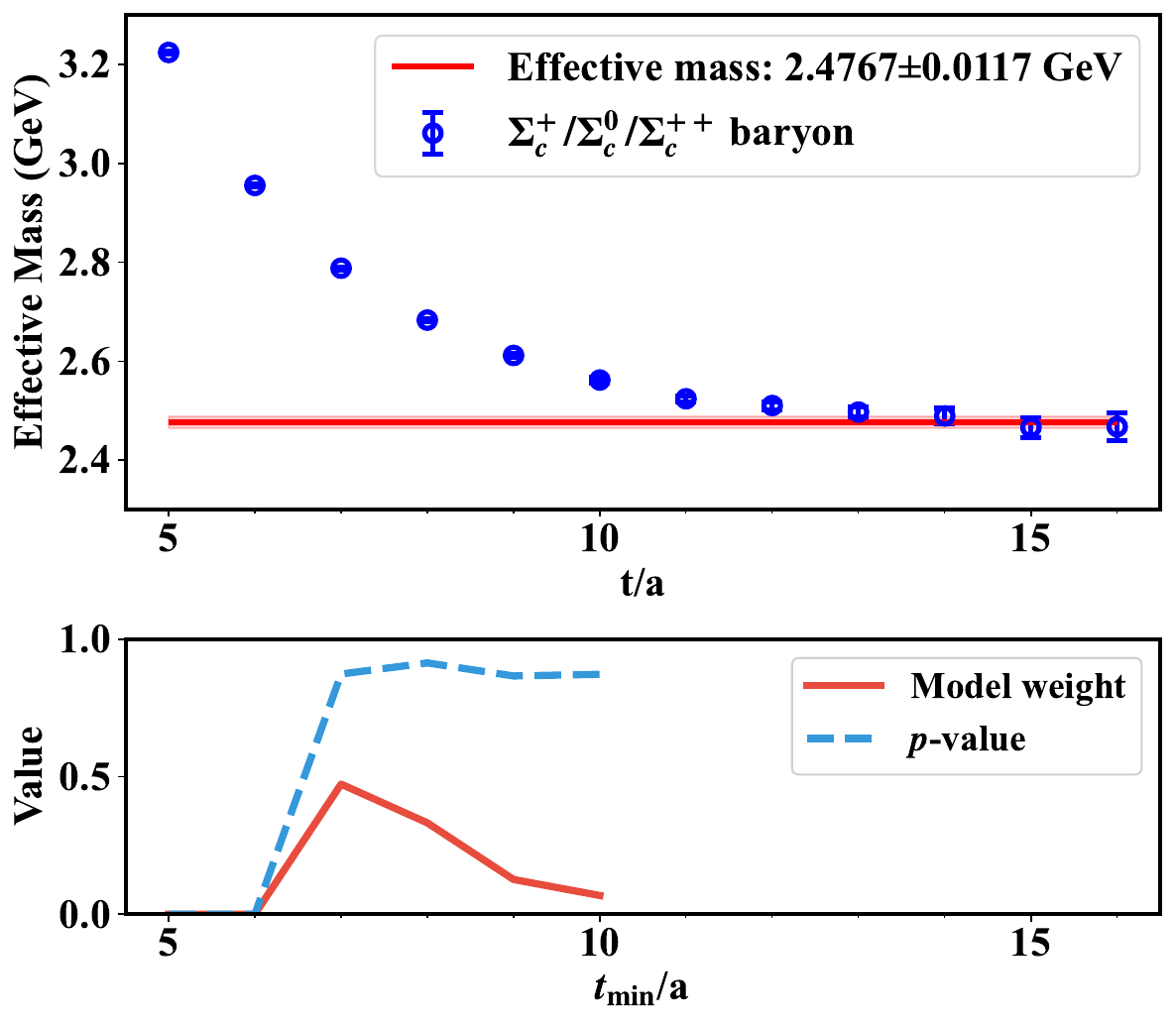}$\qquad$
\includegraphics[scale=0.25]{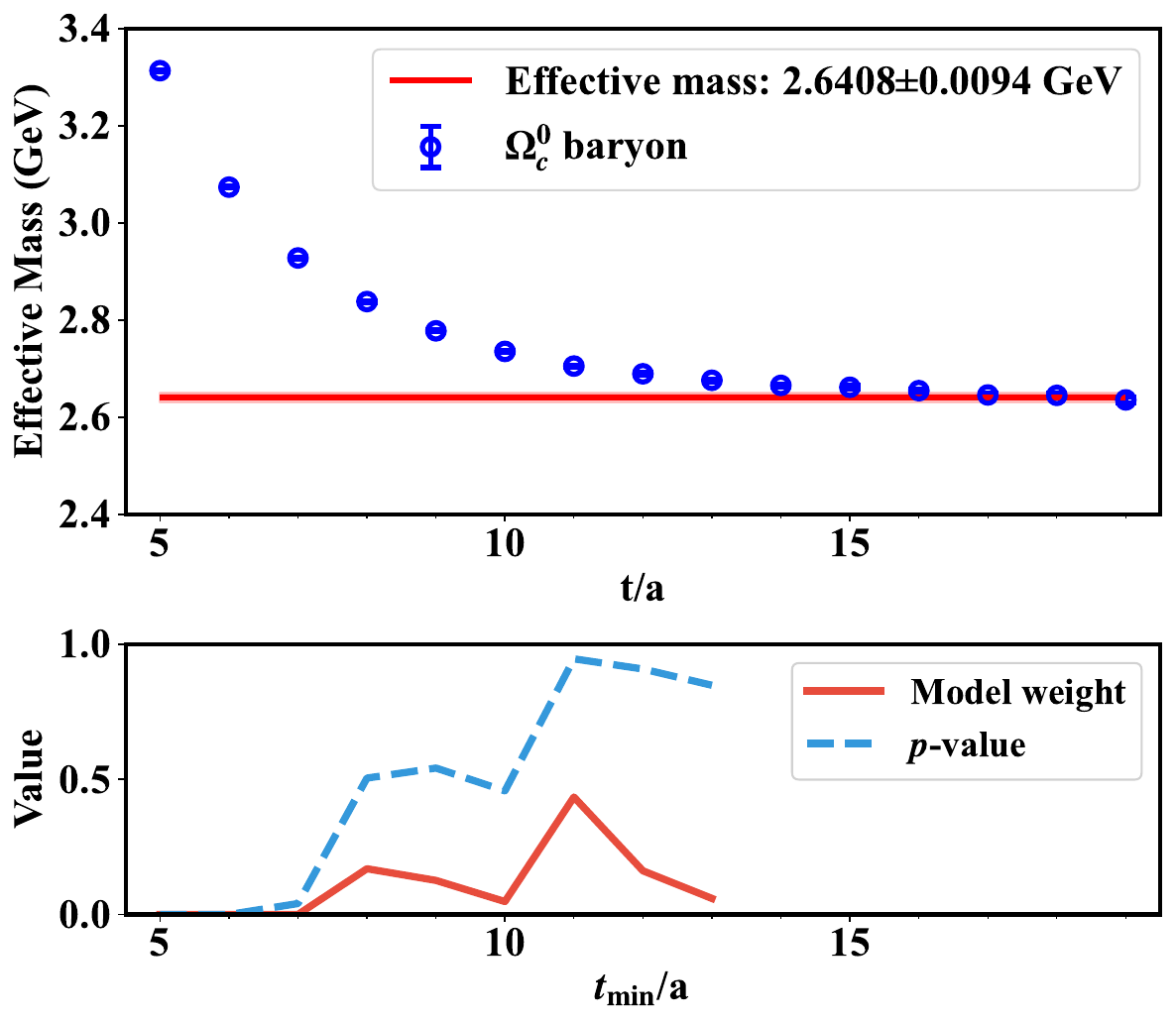}
\caption{Charmed baryons effective mass with C24P29 ensemble.}
\label{fig: C24P29}
\end{figure*}

\begin{figure*}
\centering
\includegraphics[scale=0.25]{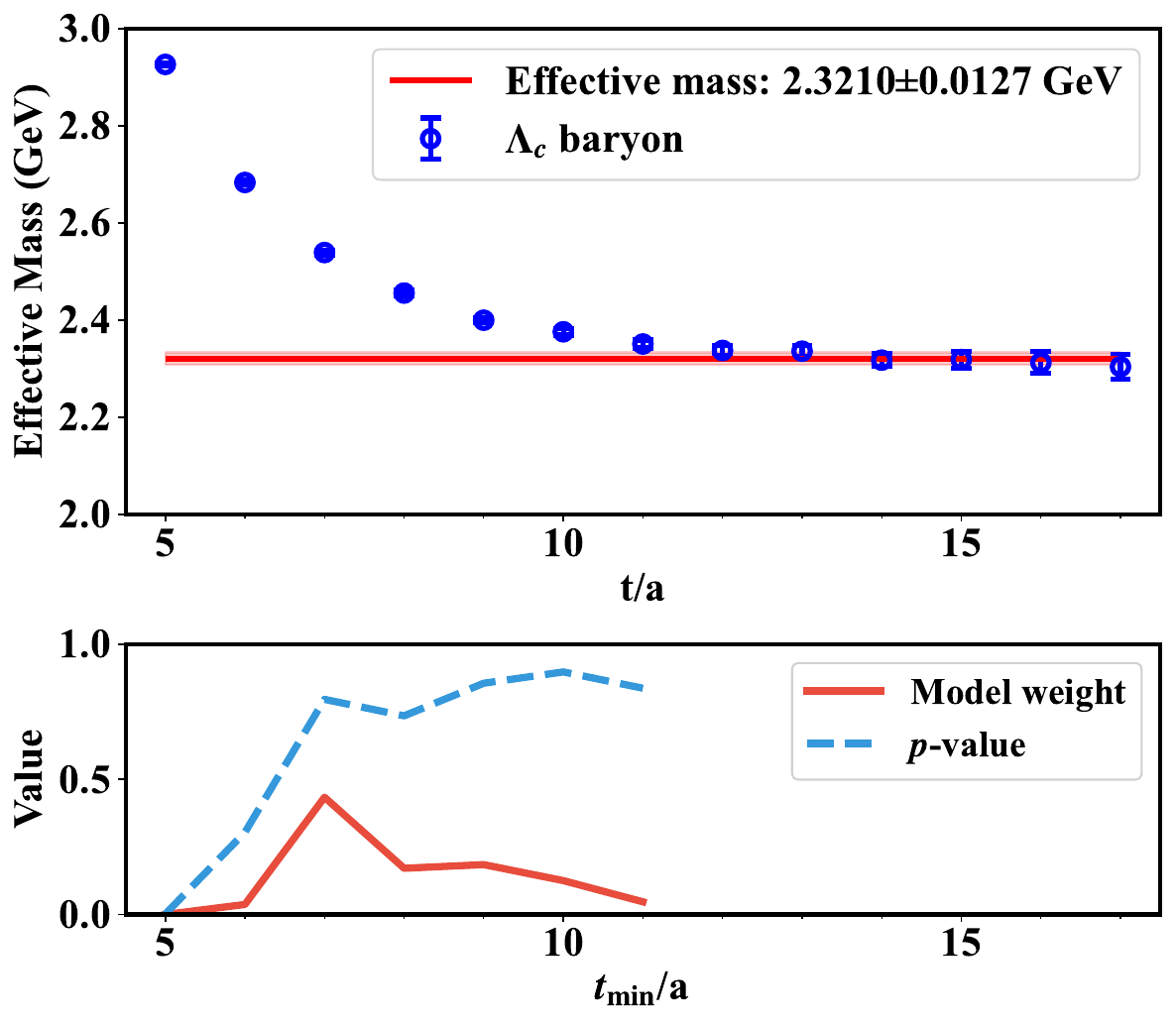}$\qquad$
\includegraphics[scale=0.25]{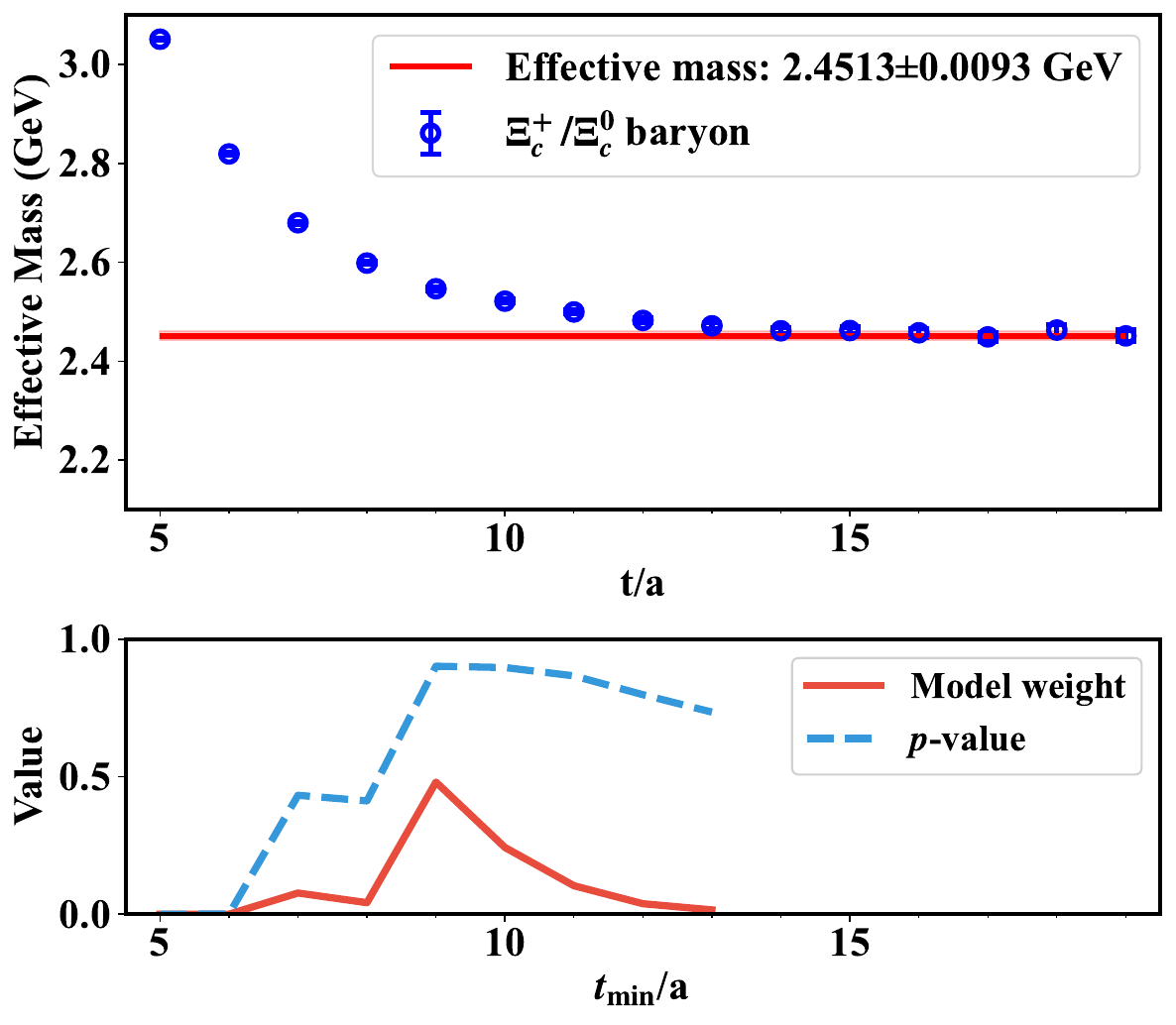}$\qquad$
\includegraphics[scale=0.25]{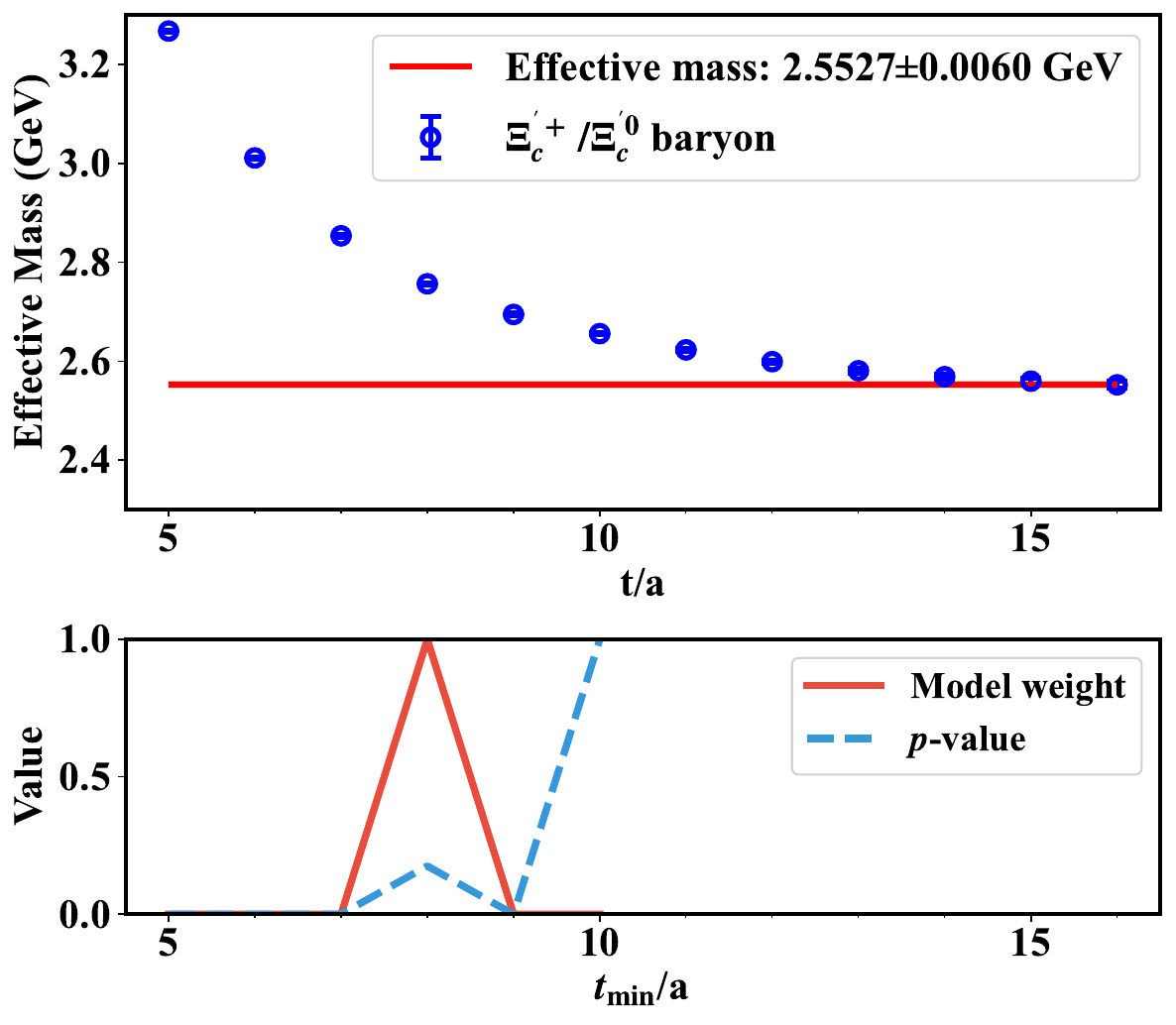}\\
\includegraphics[scale=0.25]{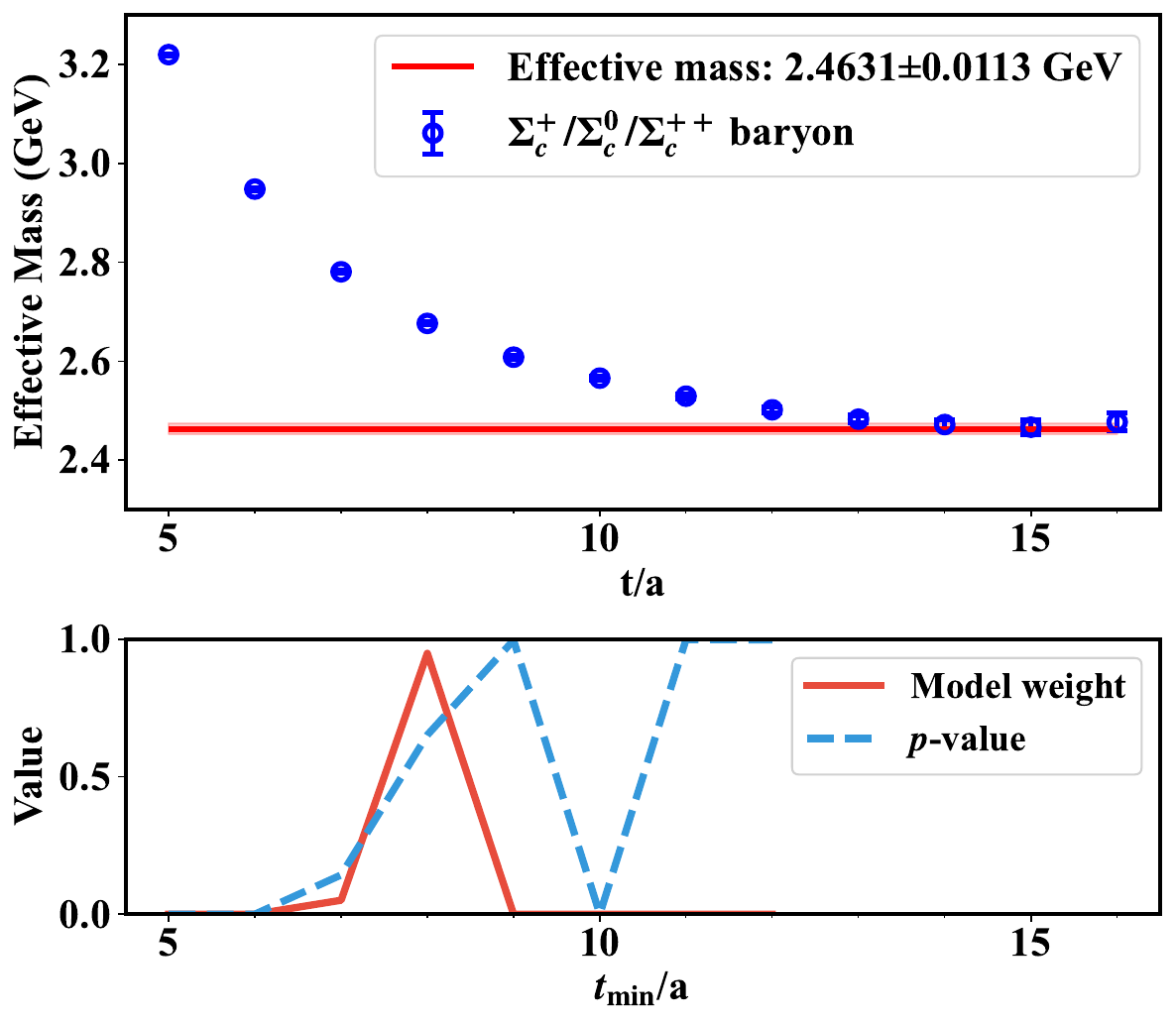}$\qquad$
\includegraphics[scale=0.25]{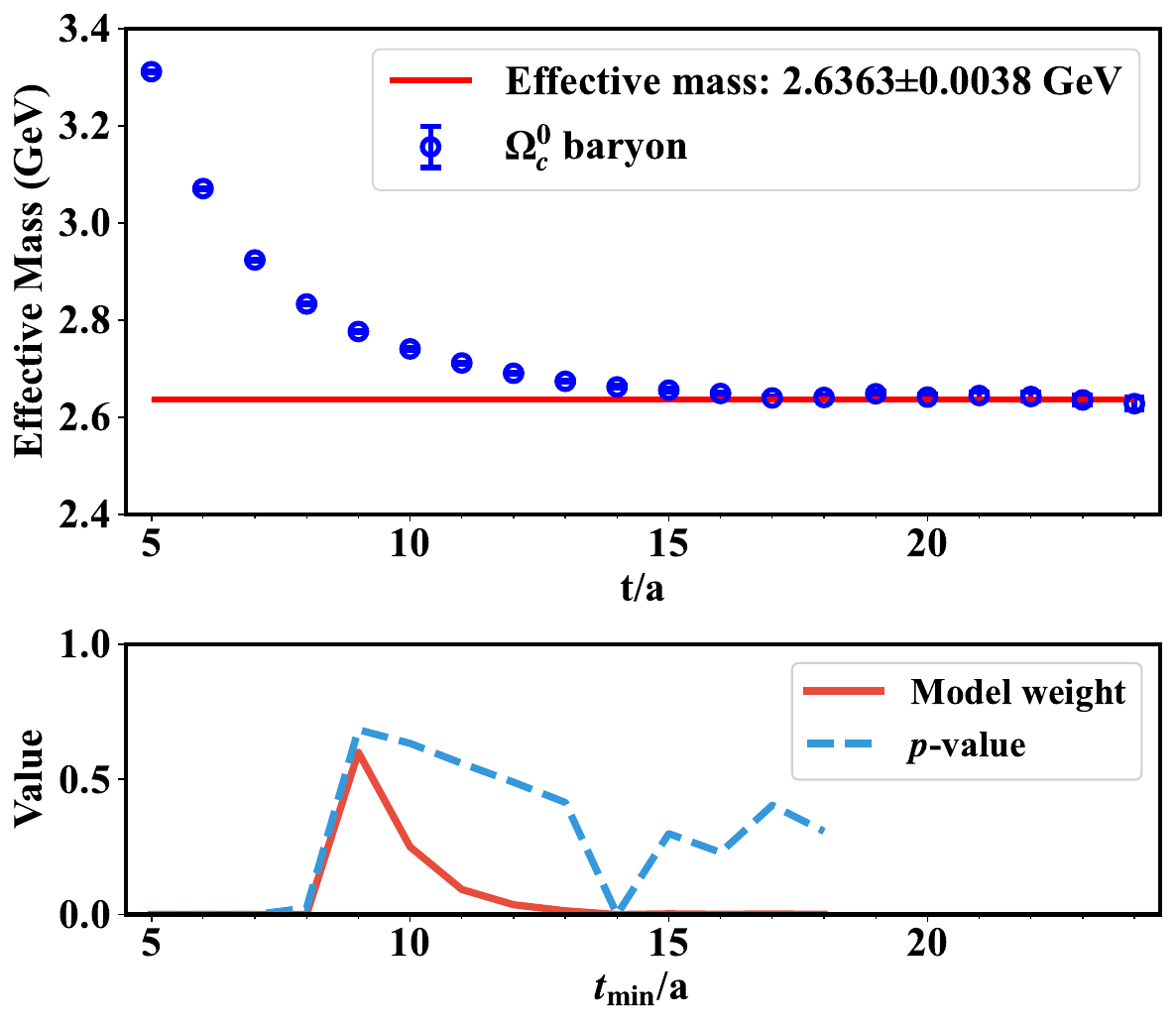}
\caption{Charmed baryons effective mass with C32P29 ensemble.}
\label{fig: C32P29}
\end{figure*}

\begin{figure*}
\centering
\includegraphics[scale=0.25]{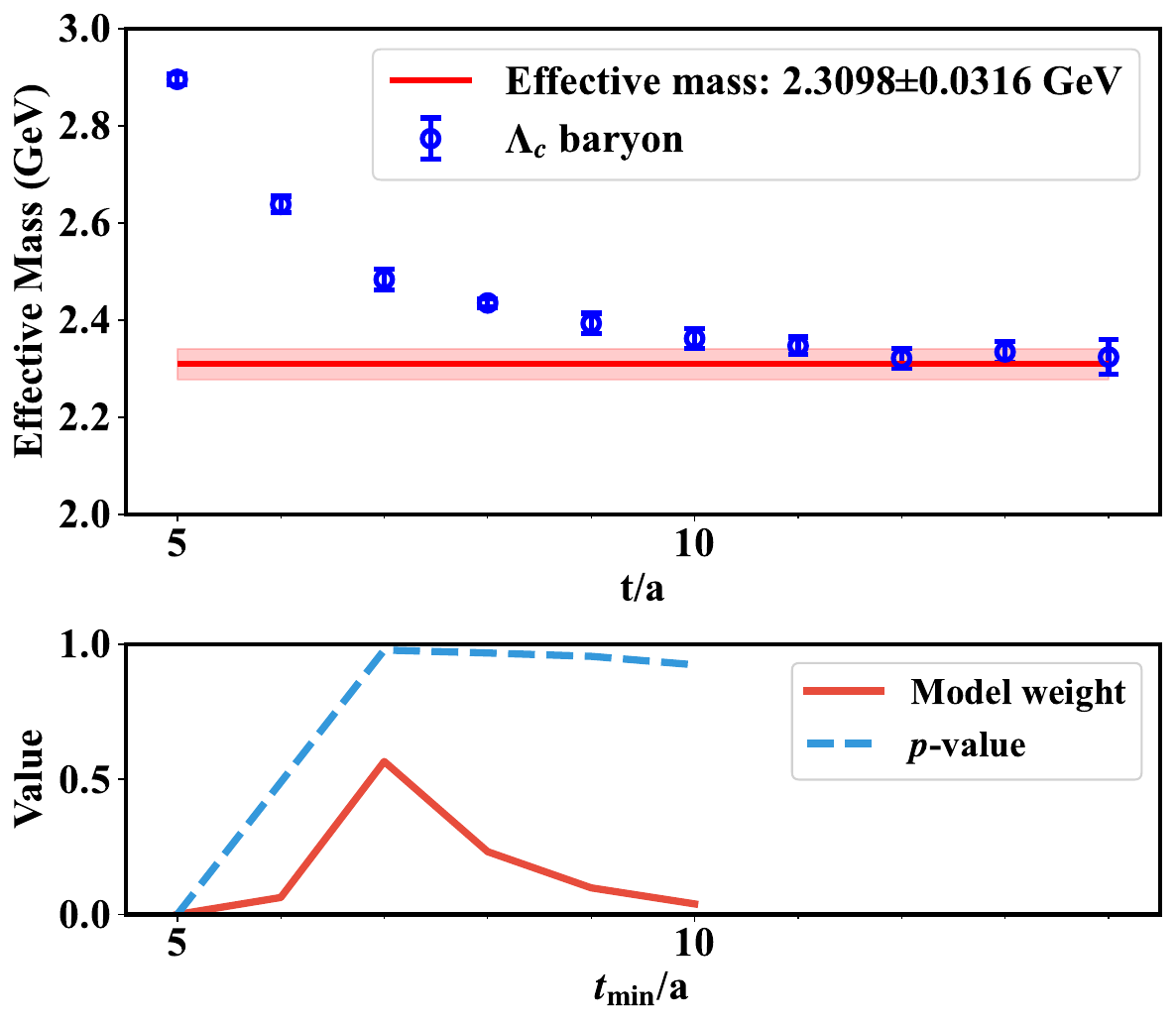}$\qquad$
\includegraphics[scale=0.25]{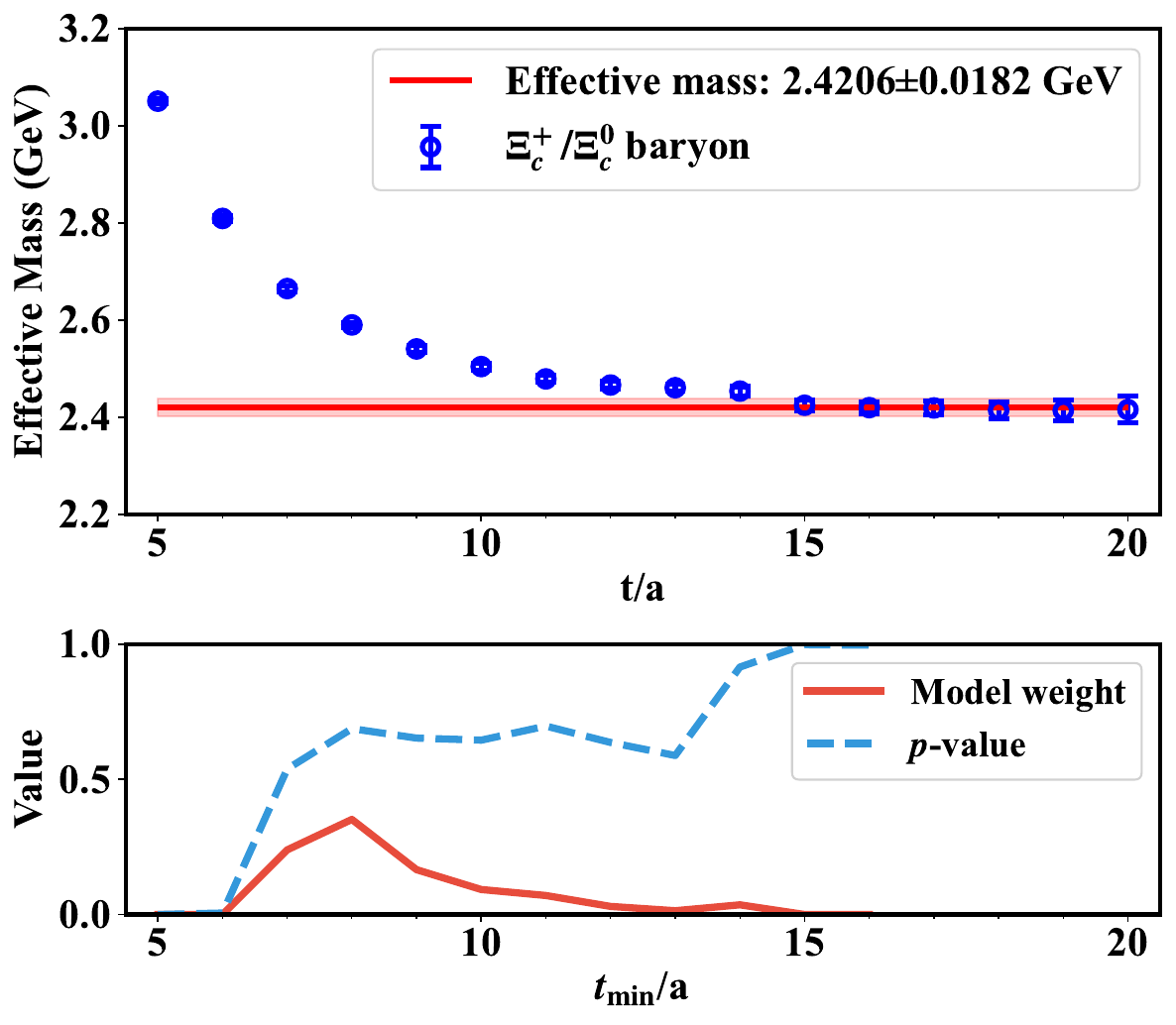}$\qquad$
\includegraphics[scale=0.25]{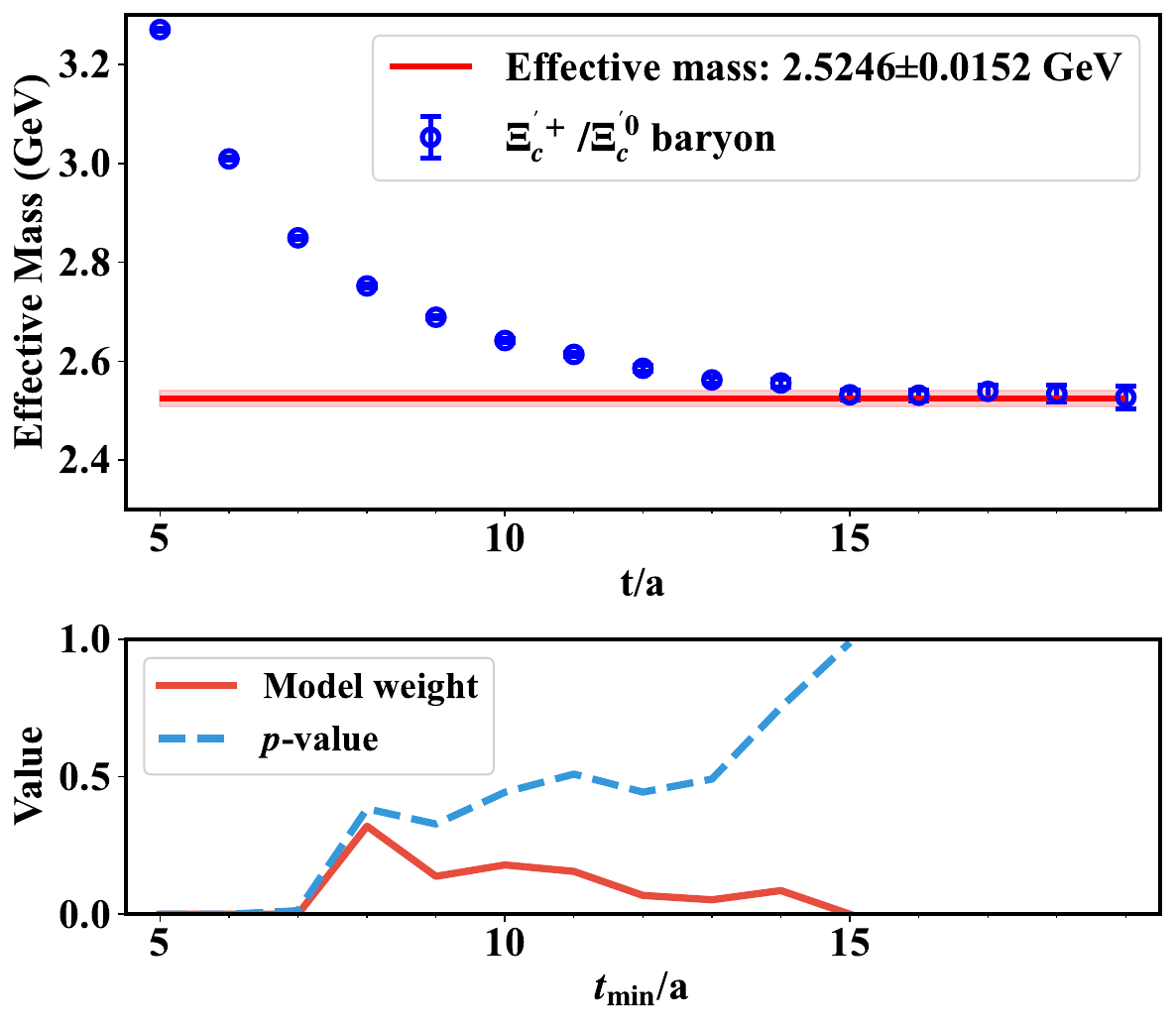}\\
\includegraphics[scale=0.25]{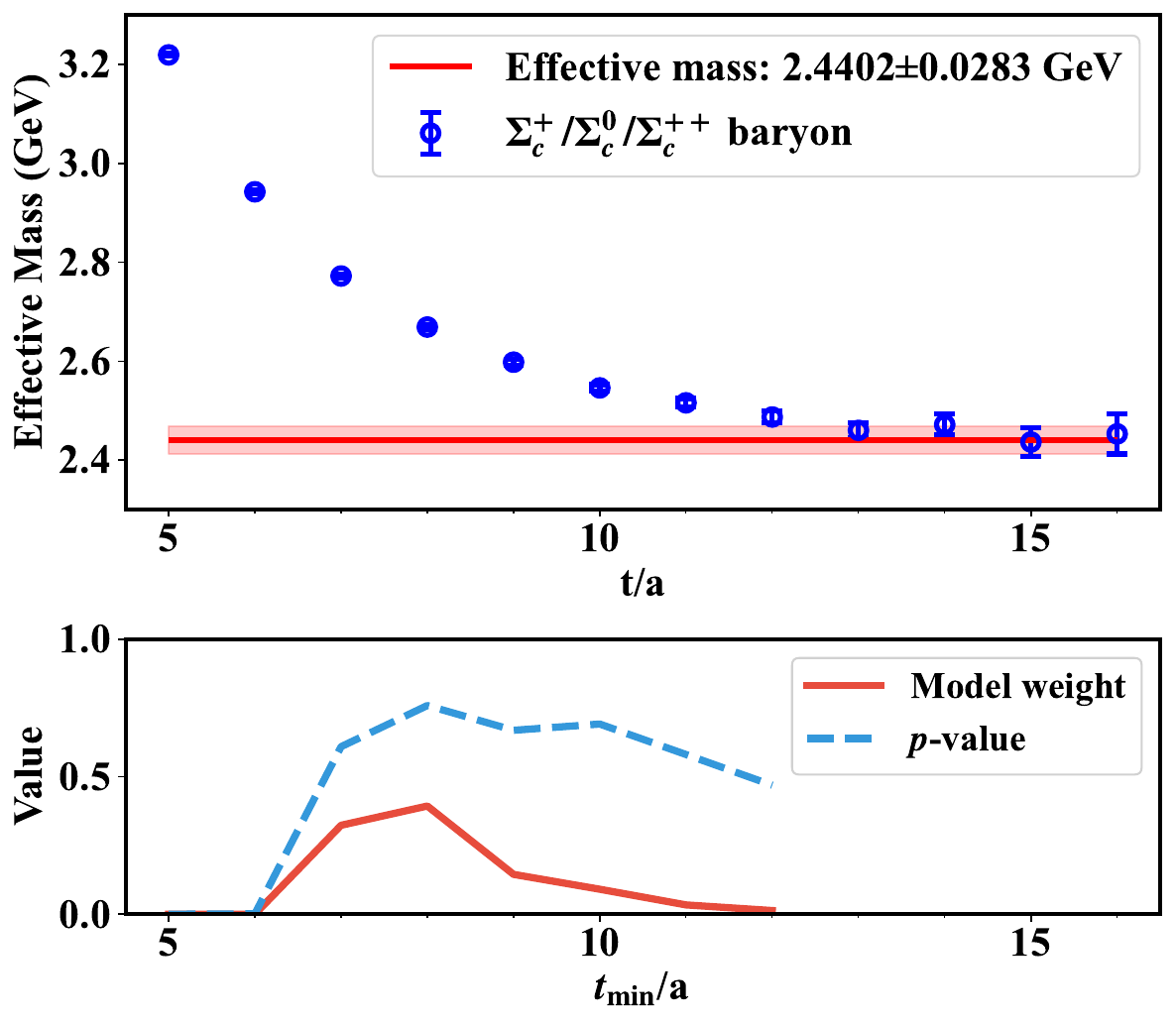}$\qquad$
\includegraphics[scale=0.25]{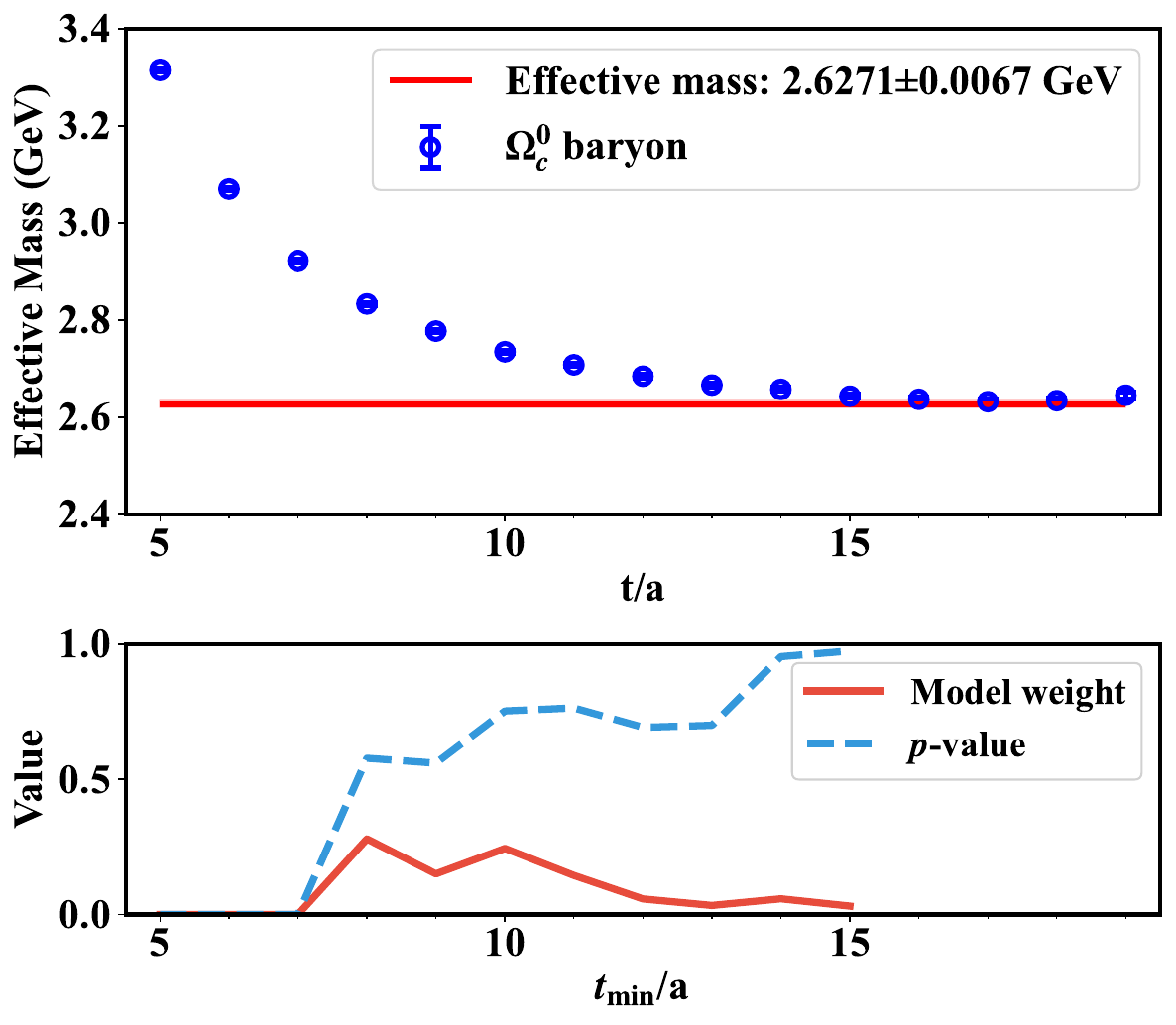}
\caption{Charm baryons effective mass  with C32P23 ensemble.}
\label{fig: C32P23}
\end{figure*}

\begin{figure*}
\centering
\includegraphics[scale=0.25]{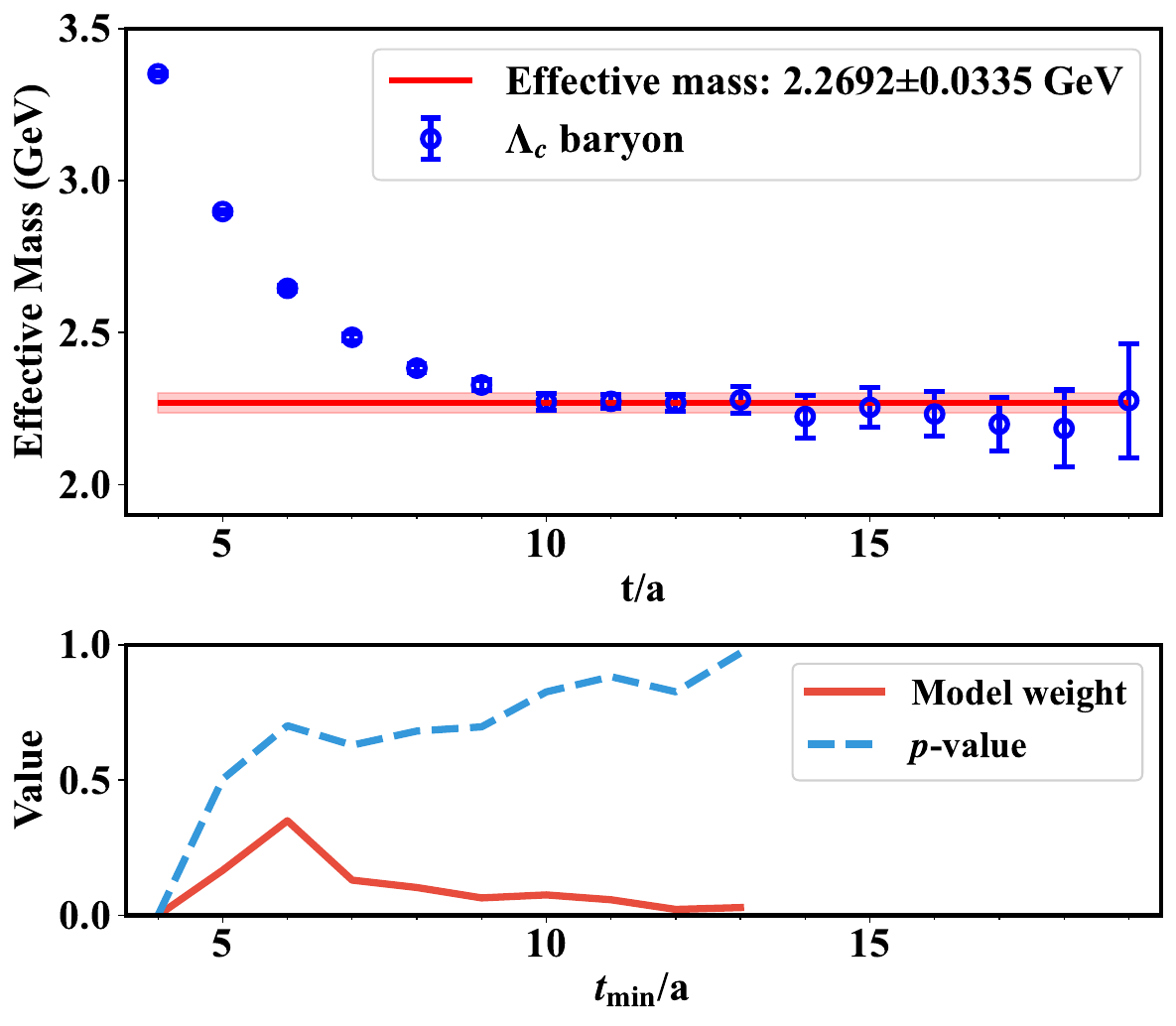}$\qquad$
\includegraphics[scale=0.25]{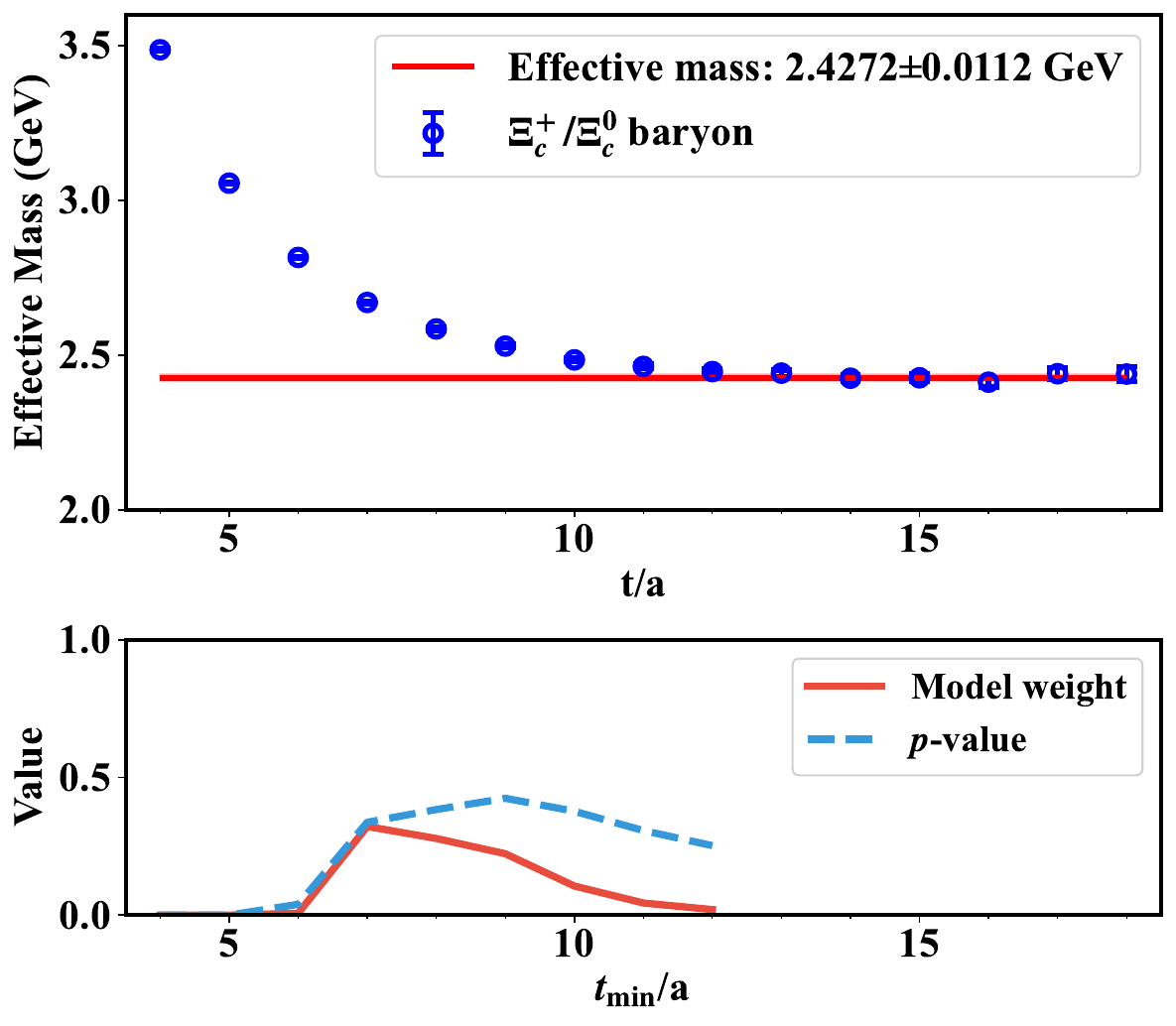}$\qquad$
\includegraphics[scale=0.25]{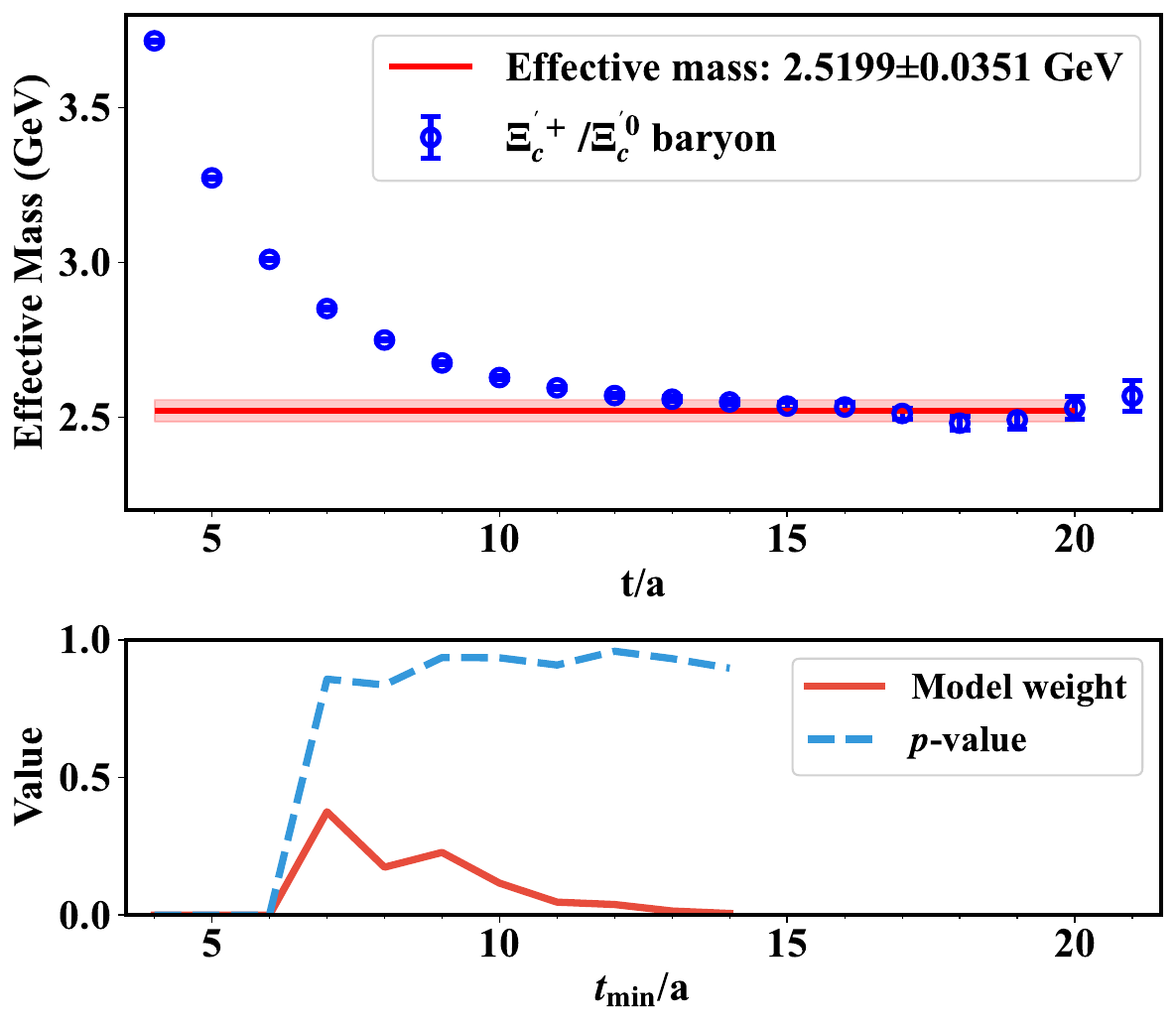}\\
\includegraphics[scale=0.25]{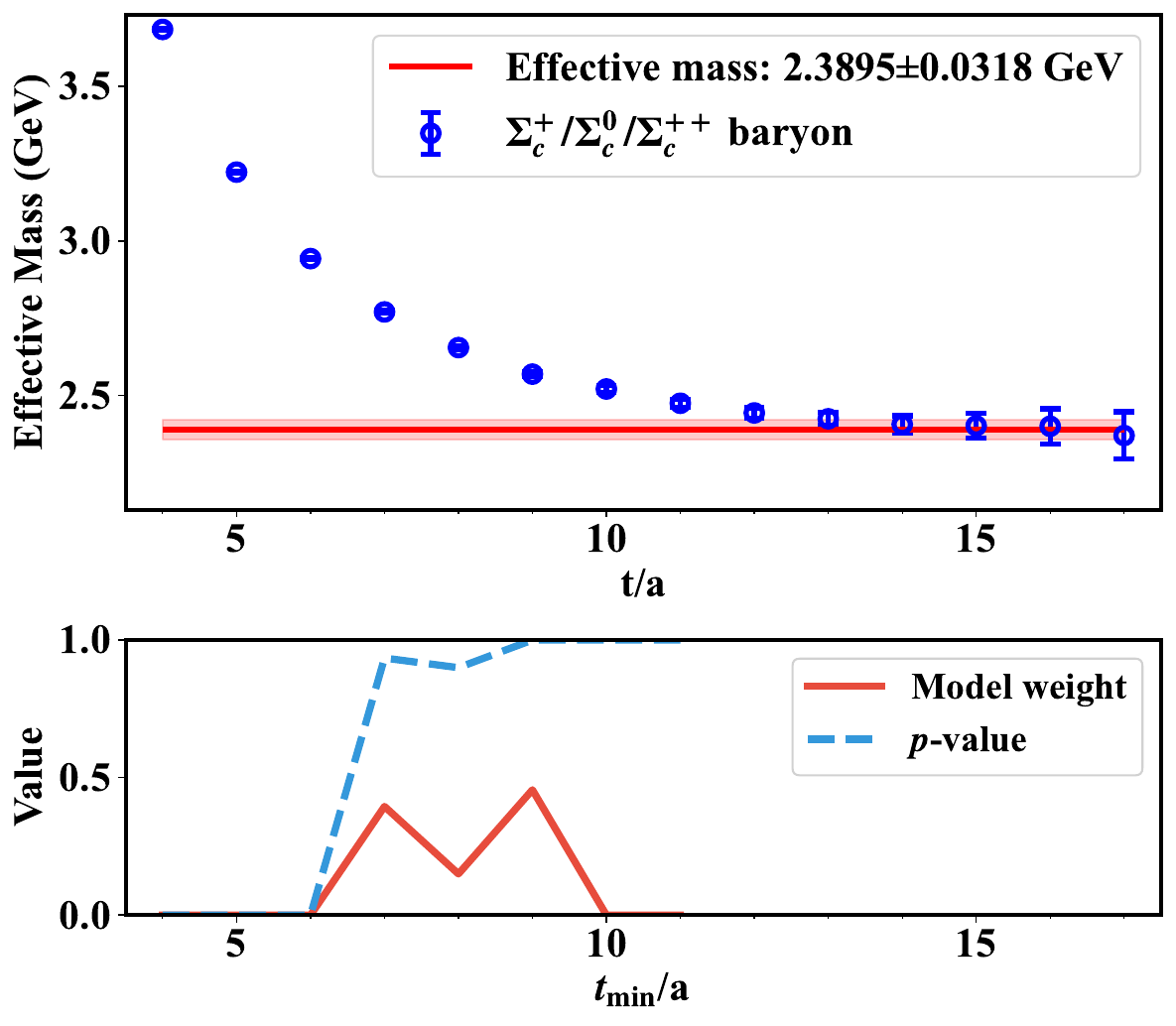}$\qquad$
\includegraphics[scale=0.25]{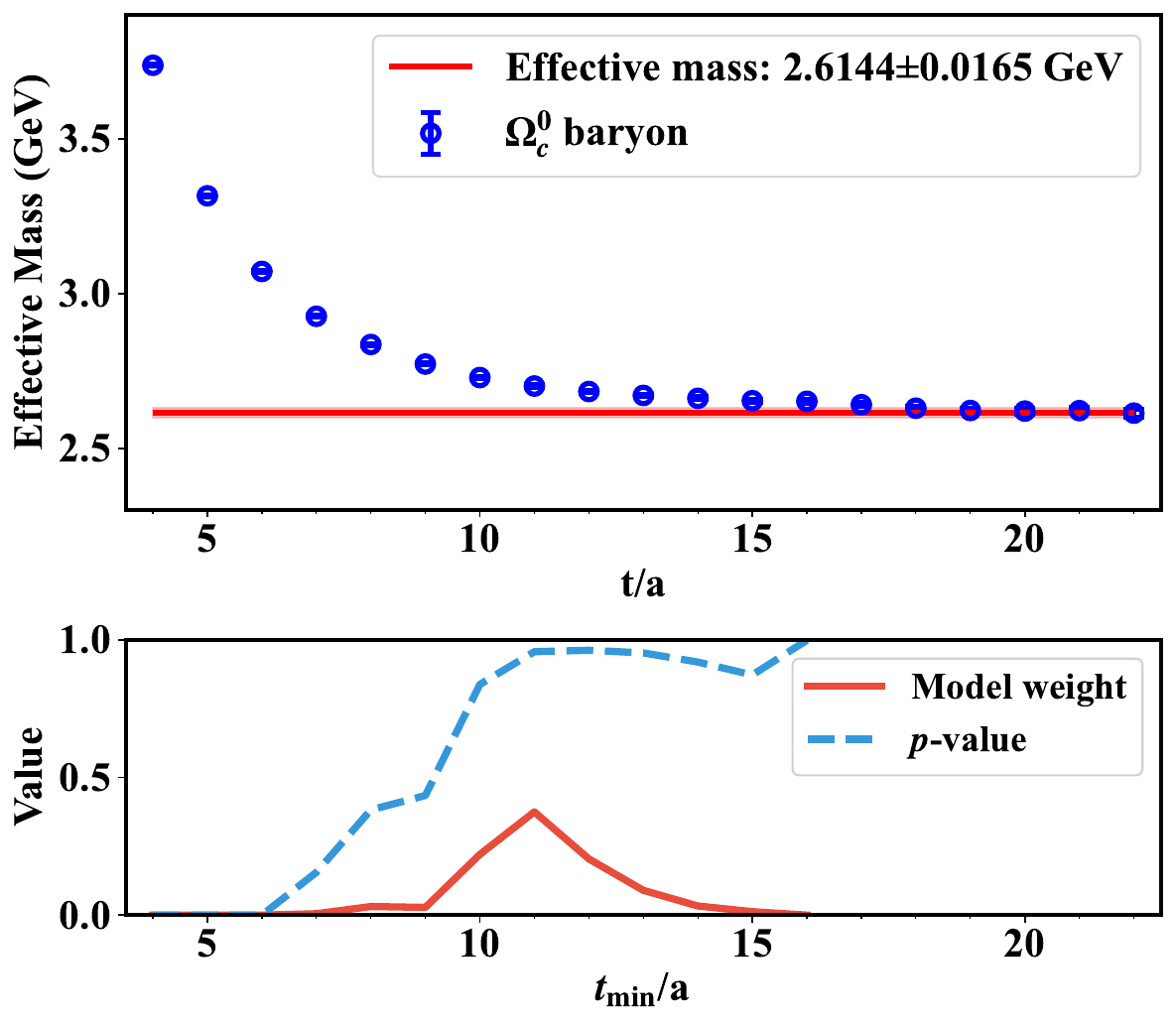}
\caption{Charmed baryons effective mass with C48P14 ensemble.}
\label{fig: C48P14}
\end{figure*}

\begin{figure*}
\centering
\includegraphics[scale=0.25]{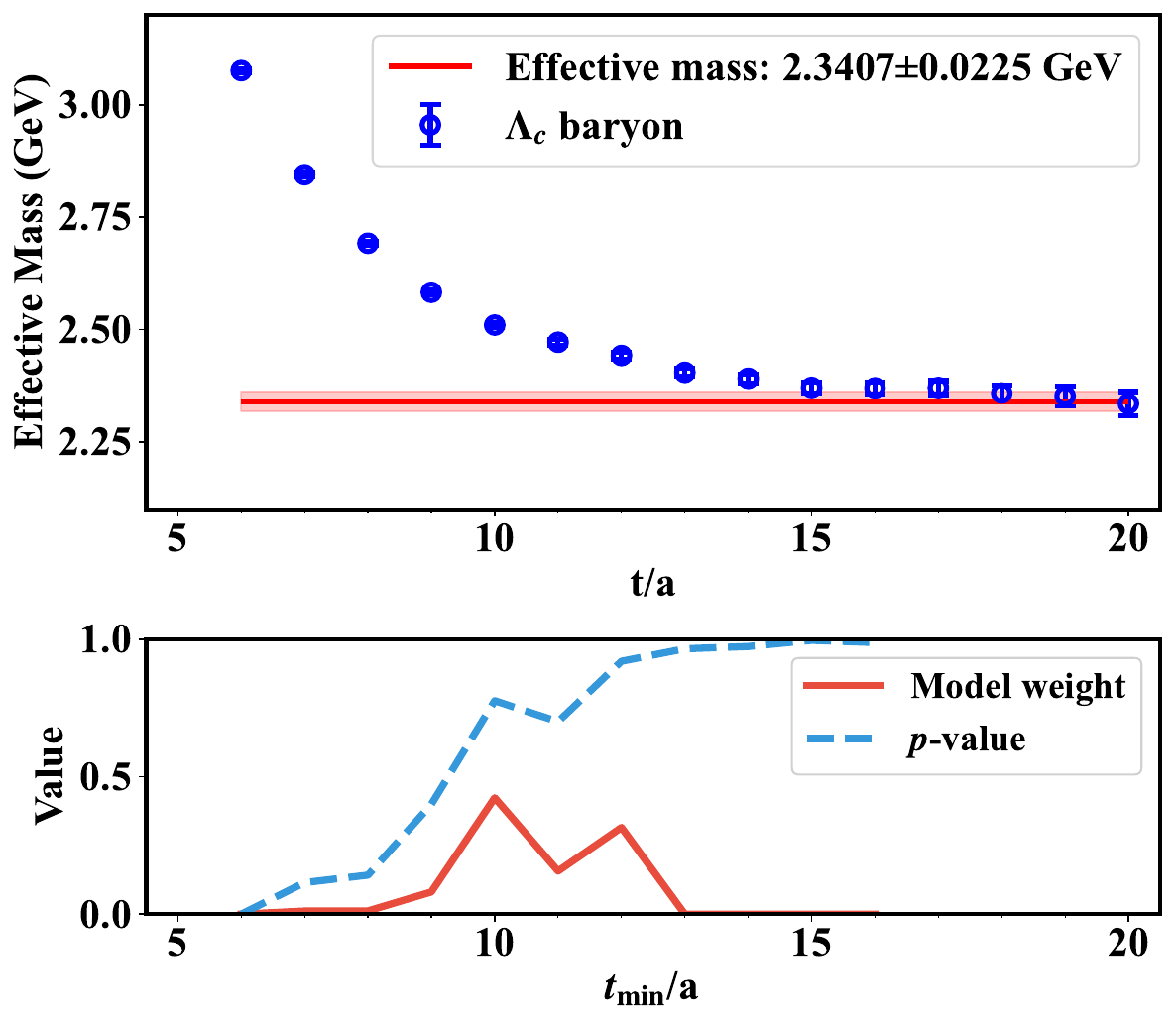}$\qquad$
\includegraphics[scale=0.25]{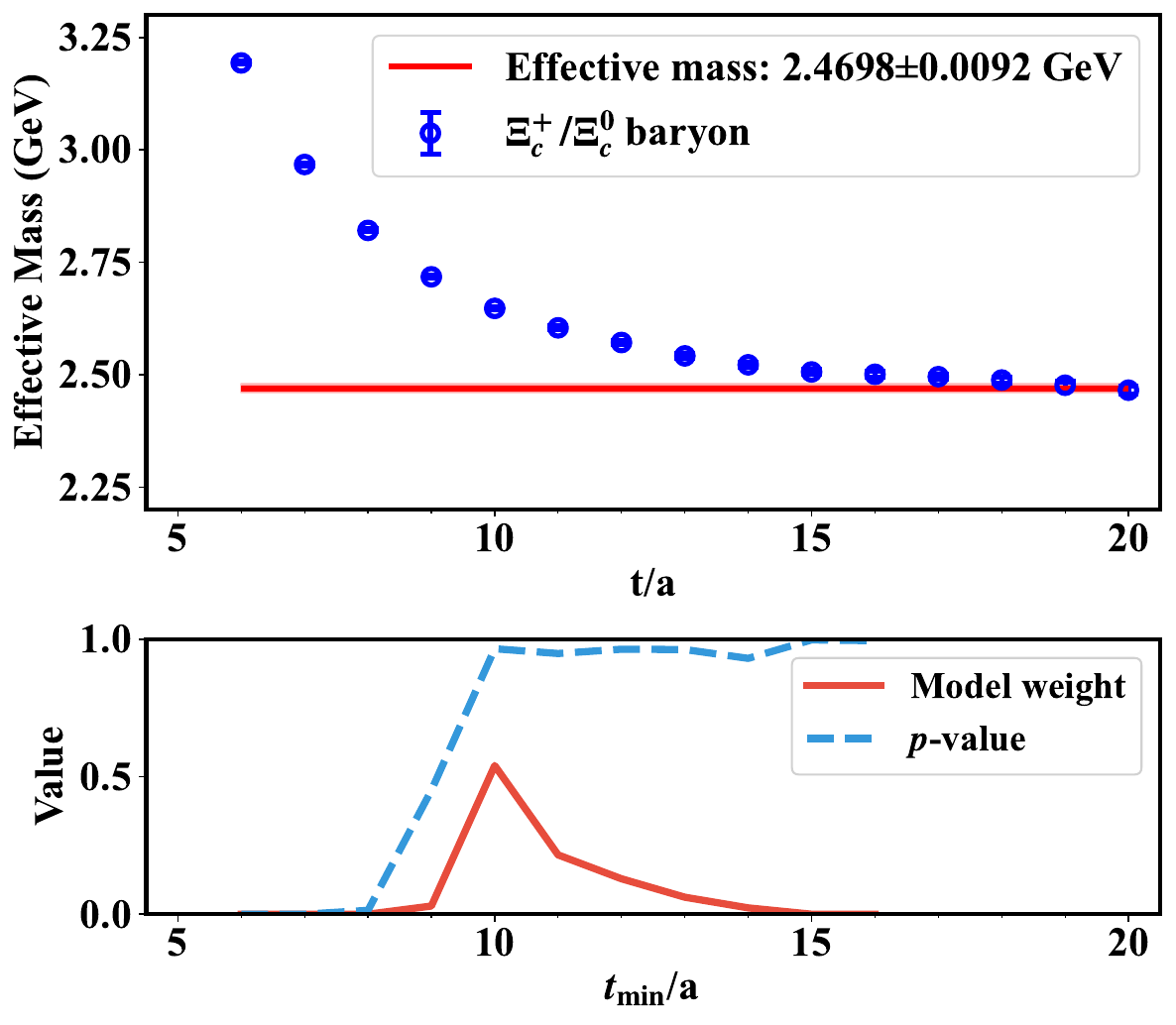}$\qquad$
\includegraphics[scale=0.25]{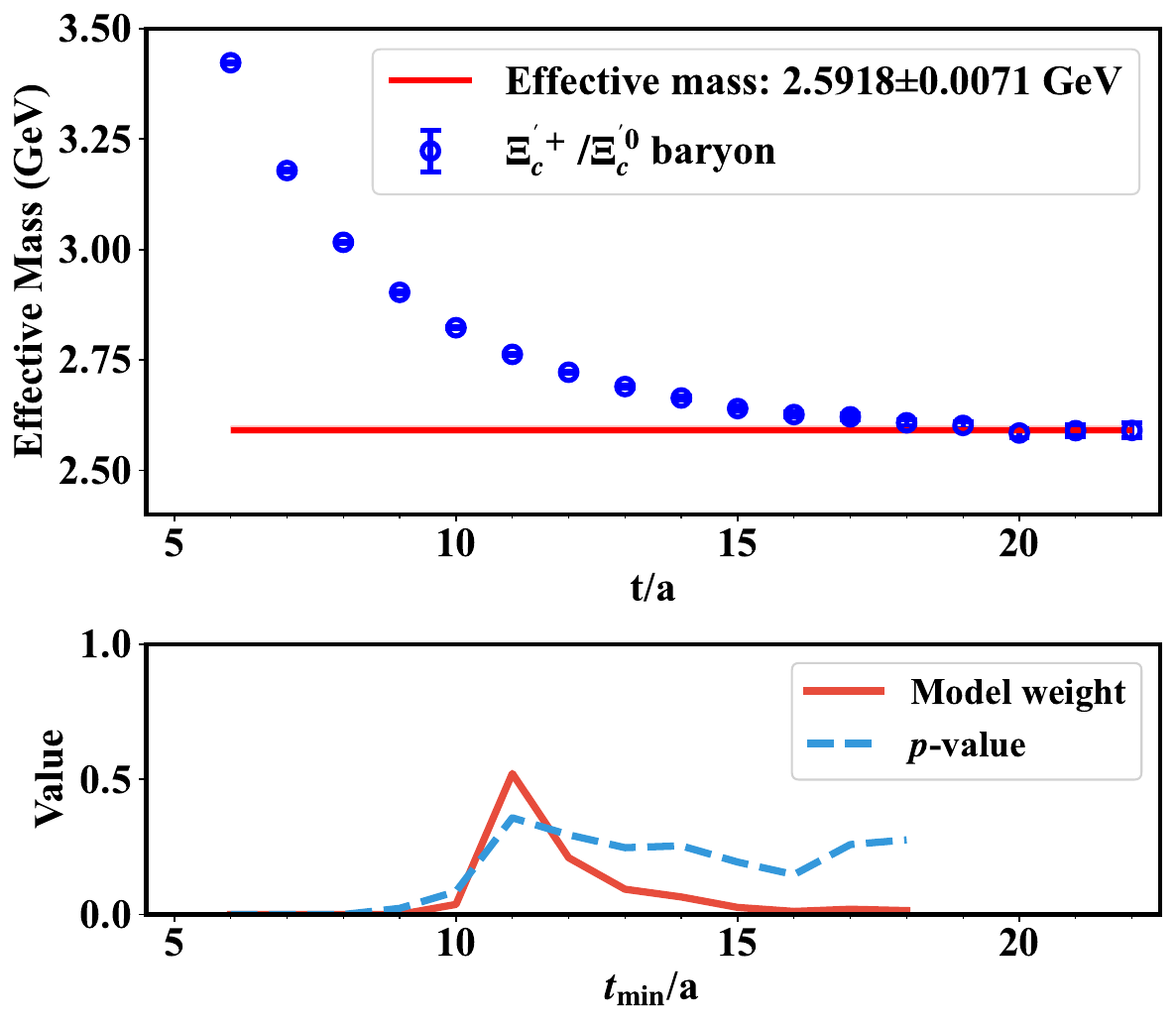}\\
\includegraphics[scale=0.25]{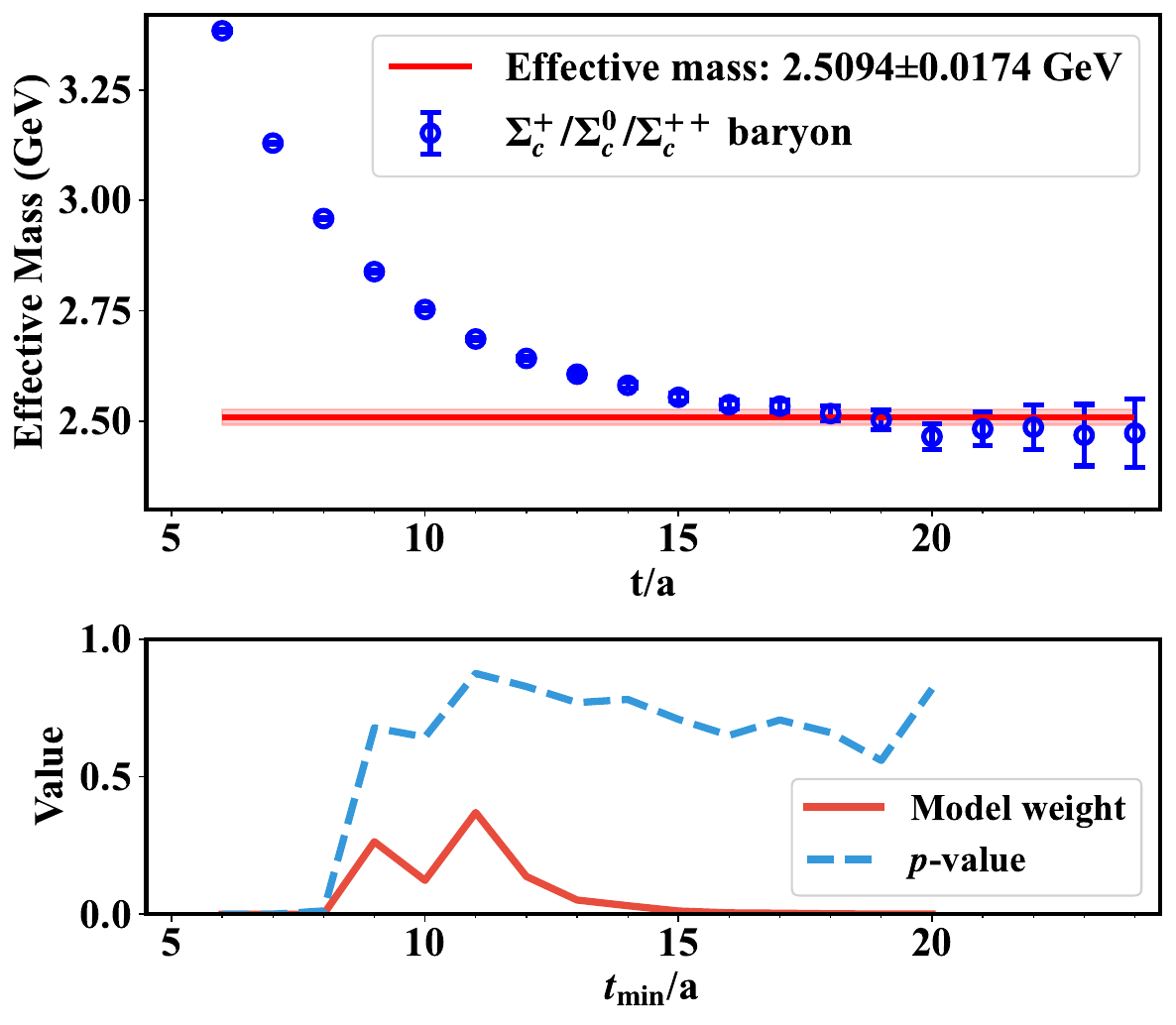}$\qquad$
\includegraphics[scale=0.25]{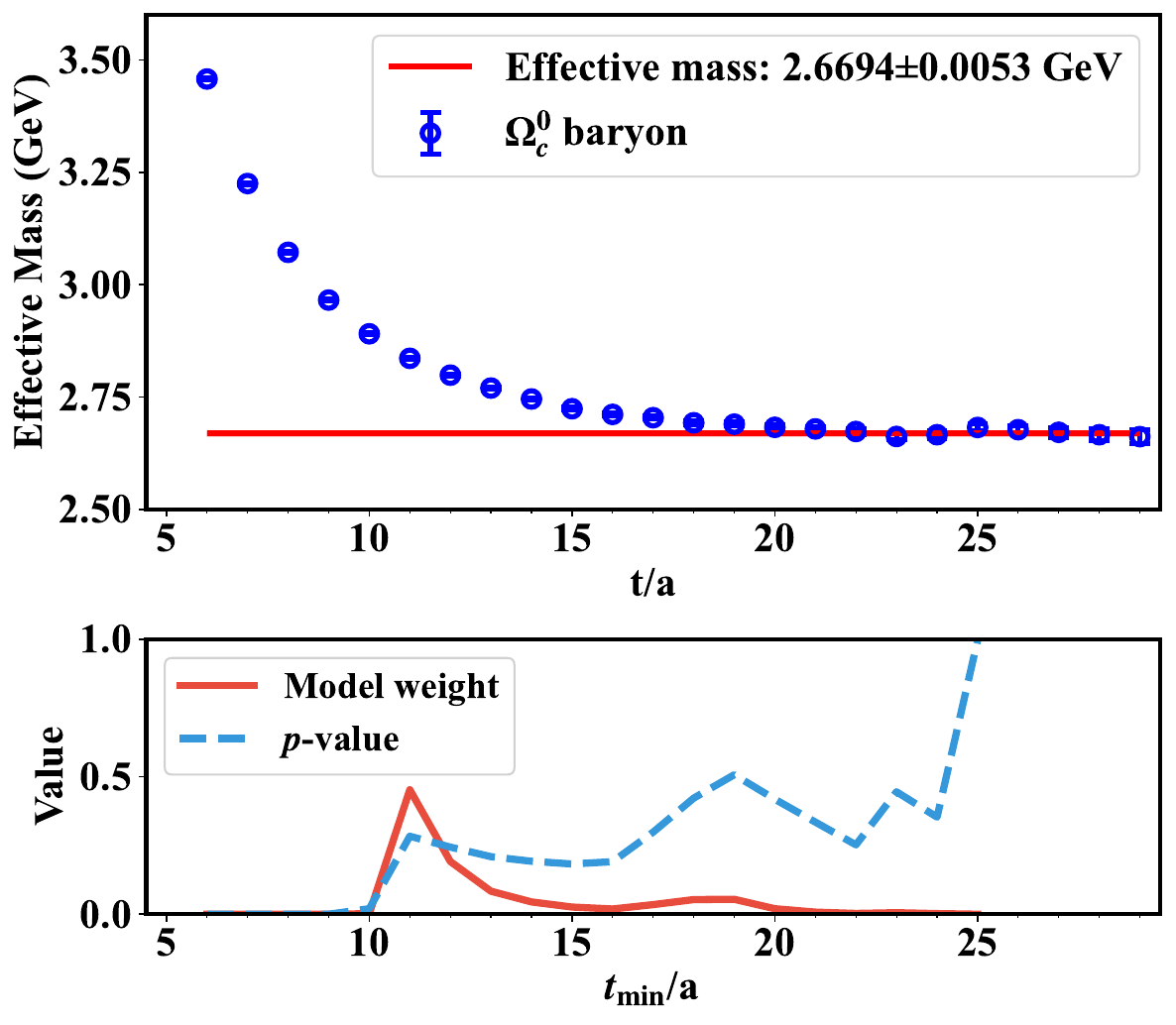}
\caption{Charmed baryons effective mass with F32P30 ensemble.}
\label{fig: F32P30}
\end{figure*}

\begin{figure*}
\centering
\includegraphics[scale=0.25]{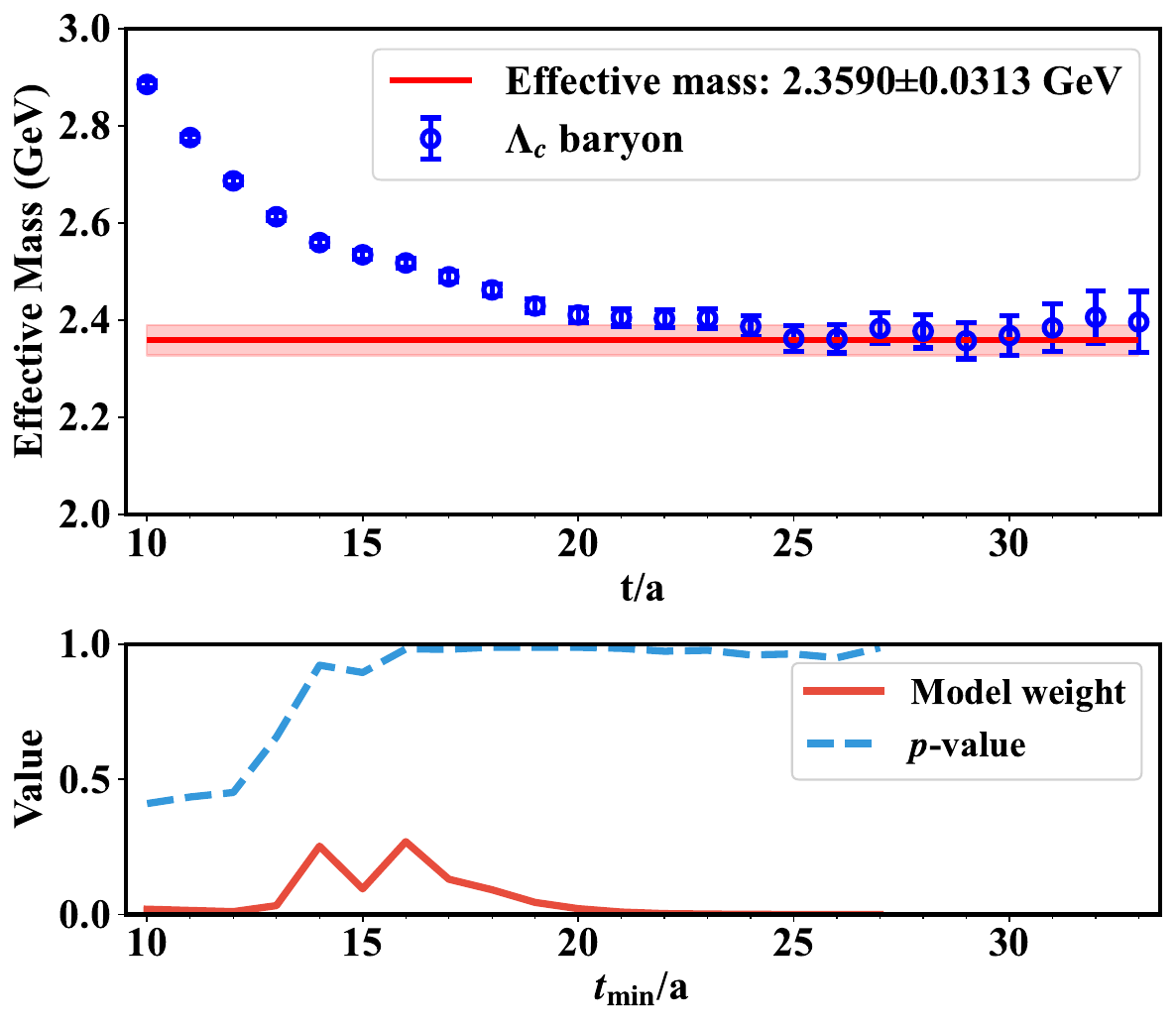}$\qquad$
\includegraphics[scale=0.25]{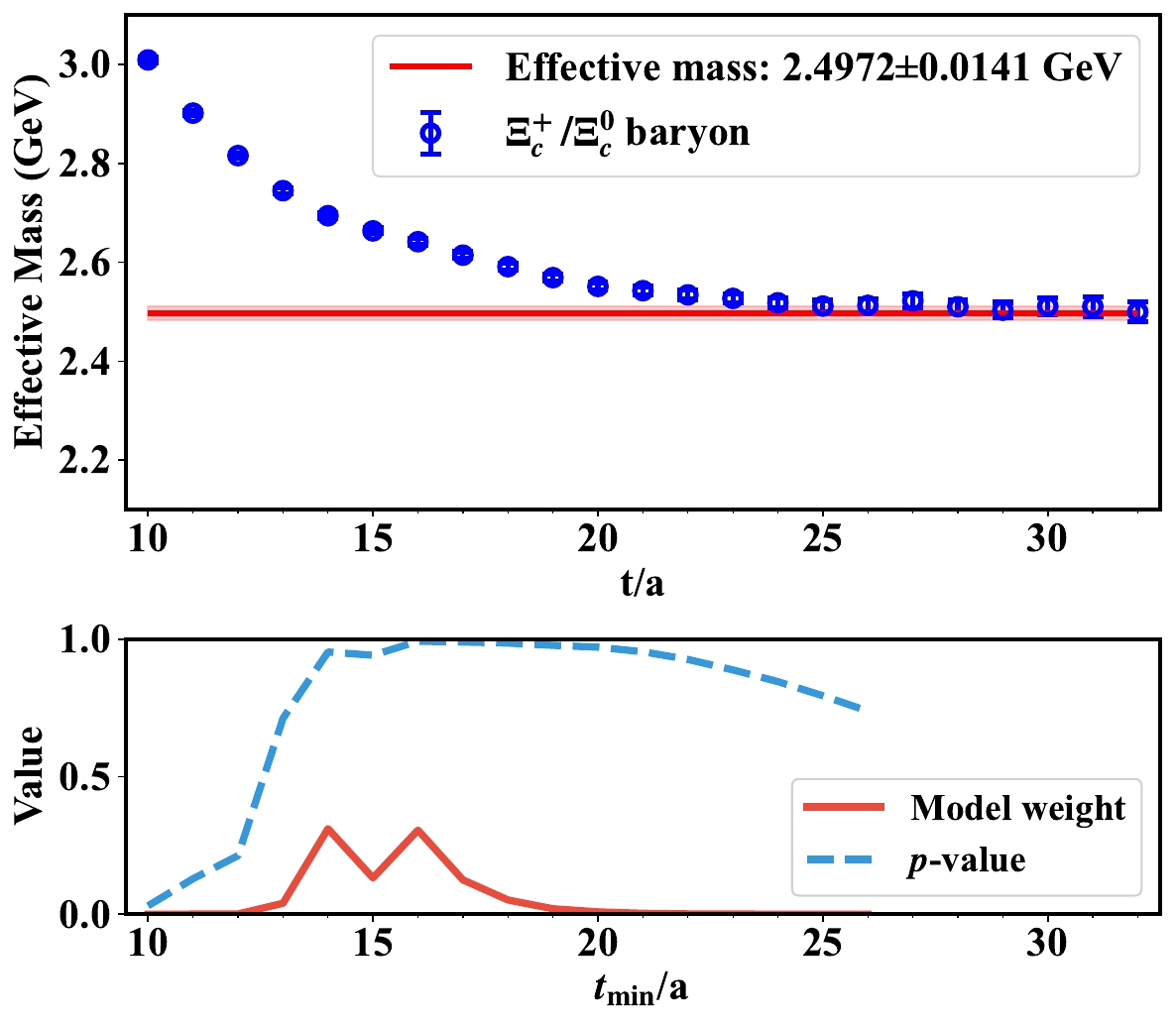}$\qquad$
\includegraphics[scale=0.25]{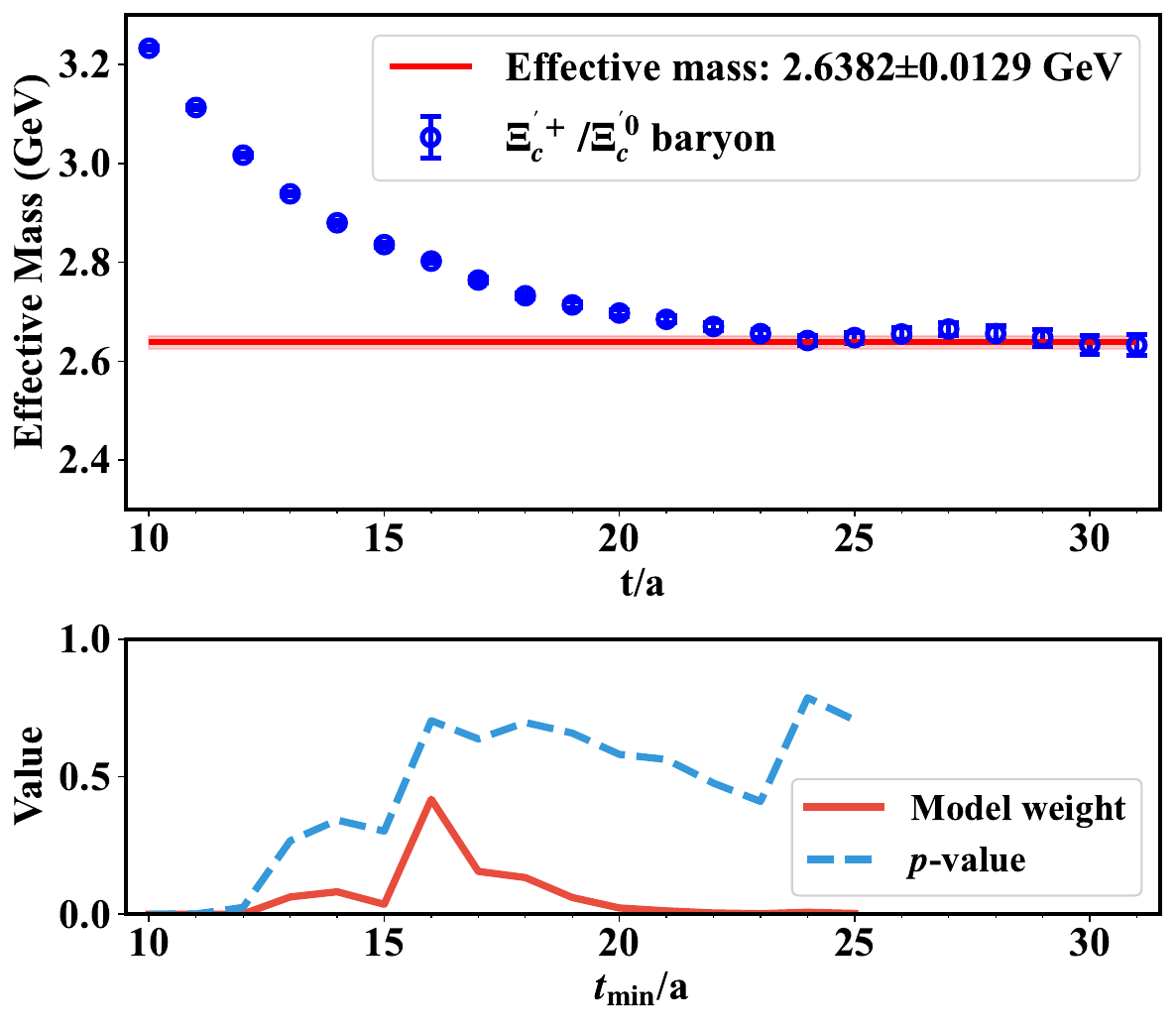}\\
\includegraphics[scale=0.25]{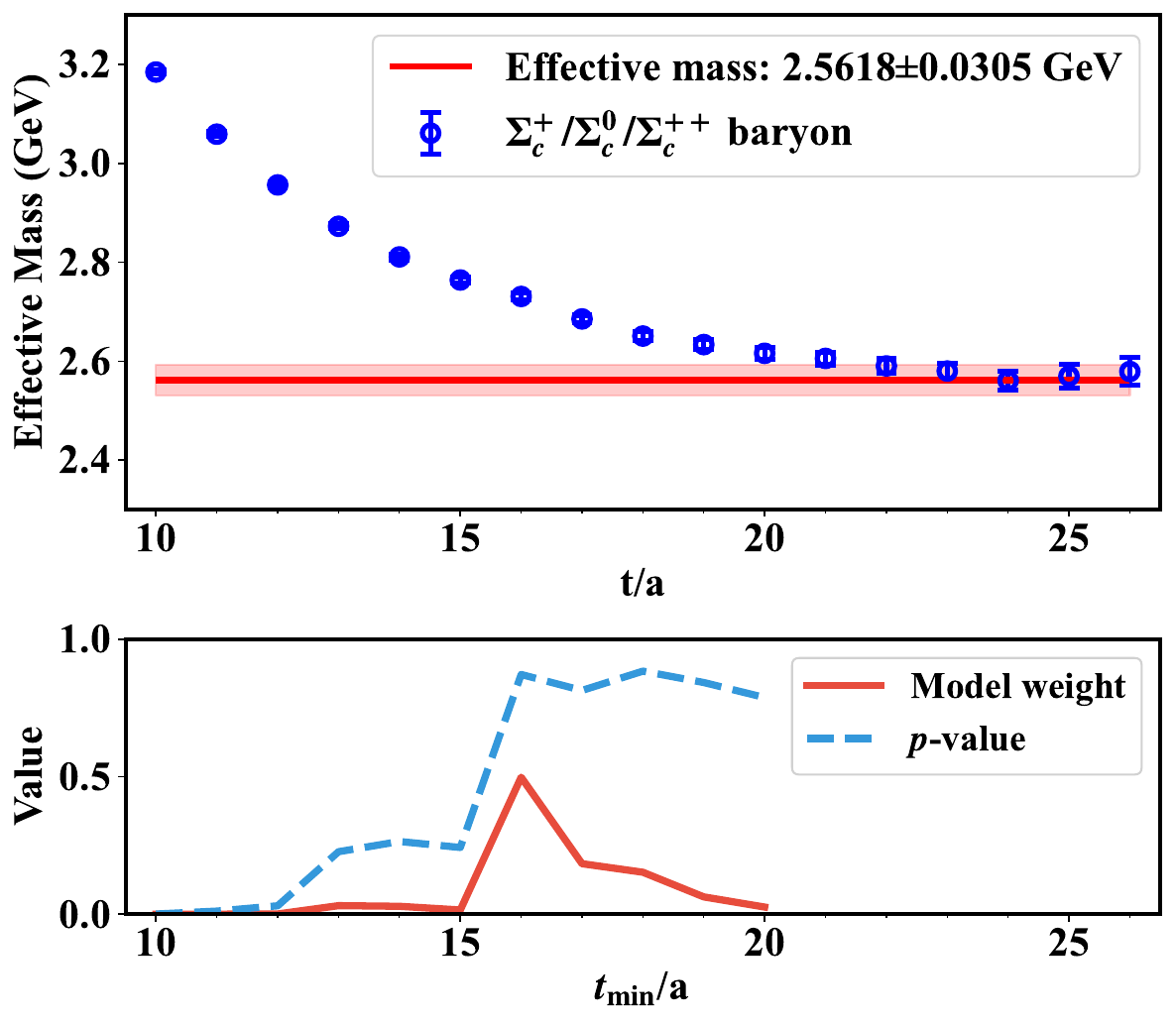}$\qquad$
\includegraphics[scale=0.25]{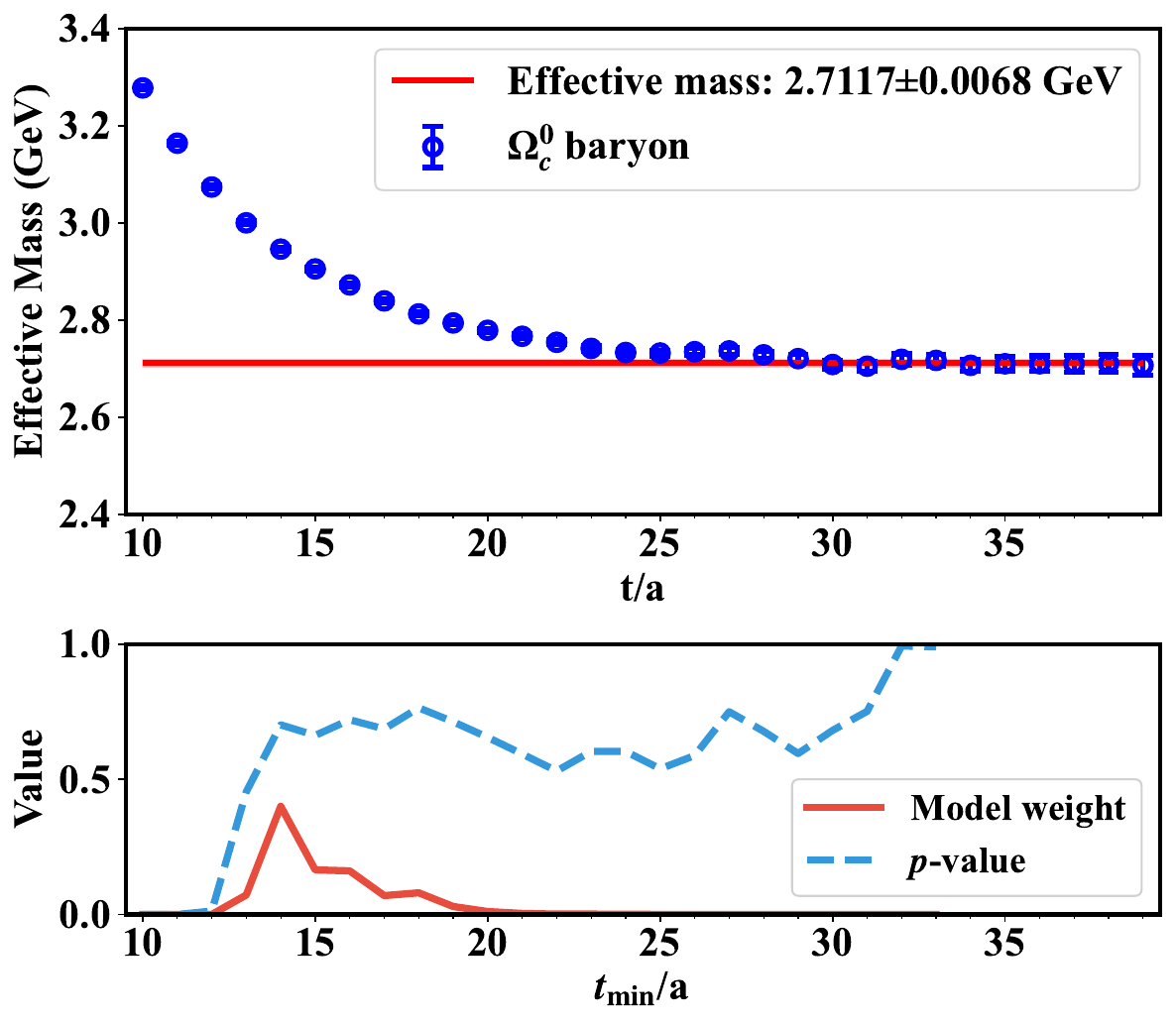}
\caption{Charmed baryons effective mass with H48P32 ensemble.}
\label{fig: H48P32}
\end{figure*}

Figs.~(\ref{fig: C24P29})-(\ref{fig: H48P32}) show the model-averaging results of the two-state fit for effective mass plateaus with different ensembles. For each baryon and ensemble, the fit plots contain two subplots: the top panel displays the effective mass data with the model-averaged plateau fit, while the bottom panel shows the standard p-value and the model weights, denoted by a blue dashed line and a red solid line, respectively.

\clearpage

\bibliographystyle{apsrev4-2}
\addcontentsline{toc}{section}{\refname}\bibliography{reference}

\end{document}